\ifdefined\pdfoutput\pdfoutput=1\fi
\documentclass[acmlarge,11pt,screen,nonacm]{acmart}
\usepackage{booktabs}
\usepackage{longtable}
\usepackage{tabularx}
\usepackage{placeins}
\setcopyright{none}
\hypersetup{keeppdfinfo=true,pdflang={en-US}}
\makeatletter
\apptocmd{\hyxmp@auto@assign@data}{%
  \hypersetup{pdfpublisher={},pdfvolumenum={},pdfissuenum={},pdfpubtype={}}%
}{}{\PackageError{article-metadata}{Metadata cleanup patch failed}
    {Check the hyperxmp metadata assignment routine.}}
\makeatother
\newcommand{\val}[4]{\ifcsname Num:#1:#2:#3:#4\endcsname\csname Num:#1:#2:#3:#4\endcsname\else\PackageError{article}{Missing generated value: #1/#2/#3/#4}{Regenerate numbers.tex and check the requested field.}\fi}
\newcommand{\diag}[2]{\ifcsname Diag:#1:#2\endcsname\csname Diag:#1:#2\endcsname\else\PackageError{article}{Missing diagnostic value: #1/#2}{Regenerate numbers.tex and check the requested field.}\fi}
\newcommand{\param}[1]{\ifcsname Param:#1\endcsname\csname Param:#1\endcsname\else\PackageError{article}{Missing generated parameter: #1}{Regenerate numbers.tex and check the parameter.}\fi}
\newcommand{\gval}[1]{\ifcsname #1\endcsname\csname #1\endcsname\else\PackageError{article}{Missing generated value: #1}{Regenerate numbers.tex and check the requested field.}\fi}

\newcommand{\prov}[3]{\gval{Provenance:#1:#2:#3}}
\newcommand{\nvd}[3]{\gval{NVD:#1:#2:#3}}
\newcommand{\stability}[3]{\gval{Stability:#1:#2:#3}}
\newcommand{\block}[3]{\gval{Block:#1:#2:#3}}
\newcommand{\early}[2]{\gval{Early:#1:#2}}
\newcommand{\mechanism}[1]{\gval{TieMechanism:Q1_2026:#1}}
\newcommand{\tie}[4]{\gval{Tie:#1:#2:#3:#4}}
\newcommand{\source}[1]{\gval{Source:#1}}
\expandafter\def\csname Param:primary_budget_pct\endcsname{5}
\expandafter\def\csname Param:lower_budget_pct\endcsname{1}
\expandafter\def\csname Param:upper_budget_pct\endcsname{10}
\expandafter\def\csname Param:outcome_days\endcsname{180}
\expandafter\def\csname Param:primary_horizon\endcsname{30}
\expandafter\def\csname Param:sensitivity_horizon\endcsname{7}
\expandafter\def\csname Source:epss_snapshots_2025\endcsname{313}
\expandafter\def\csname Source:epss_snapshots_Q1_used\endcsname{111}
\expandafter\def\csname Source:epss_model\endcsname{v2025.03.14}
\expandafter\def\csname Source:kev_release_2025\endcsname{2026.07.29}
\expandafter\def\csname Source:kev_retrieved_2025\endcsname{2 August 2026}
\expandafter\def\csname Source:kev_entries_2025\endcsname{1,656}
\expandafter\def\csname Source:kev_release_Q1\endcsname{2026.11.04}
\expandafter\def\csname Source:kev_retrieved_Q1\endcsname{5 November 2026}
\expandafter\def\csname Source:kev_entries_Q1\endcsname{1,768}
\expandafter\def\csname Source:c4_discrepancies\endcsname{6}
\expandafter\def\csname Source:epss_api_pairs\endcsname{300}
\expandafter\def\csname Source:nvd_retrieved_2025\endcsname{2 August 2026}
\expandafter\def\csname Source:epss_snapshots_Q1_retained\endcsname{113}
\expandafter\def\csname Num:2025:D30_AVAILABILITY_B01:COMMON:N\endcsname{38,368}
\expandafter\def\csname Num:2025:D30_AVAILABILITY_B01:COMMON:k\endcsname{384}
\expandafter\def\csname Num:2025:D30_AVAILABILITY_B01:COMMON:P\endcsname{40}
\expandafter\def\csname Num:2025:D30_AVAILABILITY_B01:CVSS_OPERATIONAL:captured\endcsname{5}
\expandafter\def\csname Num:2025:D30_AVAILABILITY_B01:CVSS_OPERATIONAL:minimum\endcsname{3}
\expandafter\def\csname Num:2025:D30_AVAILABILITY_B01:CVSS_OPERATIONAL:maximum\endcsname{9}
\expandafter\def\csname Num:2025:D30_AVAILABILITY_B01:CVSS_OPERATIONAL:recall\endcsname{12.5\%}
\expandafter\def\csname Num:2025:D30_AVAILABILITY_B01:CVSS_OPERATIONAL:yield\endcsname{1.30\%}
\expandafter\def\csname Num:2025:D30_AVAILABILITY_B01:EPSS_OPERATIONAL:captured\endcsname{8}
\expandafter\def\csname Num:2025:D30_AVAILABILITY_B01:EPSS_OPERATIONAL:minimum\endcsname{8}
\expandafter\def\csname Num:2025:D30_AVAILABILITY_B01:EPSS_OPERATIONAL:maximum\endcsname{8}
\expandafter\def\csname Num:2025:D30_AVAILABILITY_B01:EPSS_OPERATIONAL:recall\endcsname{20.0\%}
\expandafter\def\csname Num:2025:D30_AVAILABILITY_B01:EPSS_OPERATIONAL:yield\endcsname{2.08\%}
\expandafter\def\csname Num:2025:D30_AVAILABILITY_B05:COMMON:N\endcsname{38,368}
\expandafter\def\csname Num:2025:D30_AVAILABILITY_B05:COMMON:k\endcsname{1,919}
\expandafter\def\csname Num:2025:D30_AVAILABILITY_B05:COMMON:P\endcsname{40}
\expandafter\def\csname Num:2025:D30_AVAILABILITY_B05:COMMON:base_rate\endcsname{0.10\%}
\expandafter\def\csname Num:2025:D30_AVAILABILITY_B05:COMMON:neither\endcsname{21}
\expandafter\def\csname Num:2025:D30_AVAILABILITY_B05:CVSS_OPERATIONAL:captured\endcsname{5}
\expandafter\def\csname Num:2025:D30_AVAILABILITY_B05:CVSS_OPERATIONAL:minimum\endcsname{3}
\expandafter\def\csname Num:2025:D30_AVAILABILITY_B05:CVSS_OPERATIONAL:maximum\endcsname{9}
\expandafter\def\csname Num:2025:D30_AVAILABILITY_B05:CVSS_OPERATIONAL:recall\endcsname{12.5\%}
\expandafter\def\csname Num:2025:D30_AVAILABILITY_B05:CVSS_OPERATIONAL:yield\endcsname{0.26\%}
\expandafter\def\csname Num:2025:D30_AVAILABILITY_B05:EPSS_OPERATIONAL:captured\endcsname{16}
\expandafter\def\csname Num:2025:D30_AVAILABILITY_B05:EPSS_OPERATIONAL:minimum\endcsname{16}
\expandafter\def\csname Num:2025:D30_AVAILABILITY_B05:EPSS_OPERATIONAL:maximum\endcsname{16}
\expandafter\def\csname Num:2025:D30_AVAILABILITY_B05:EPSS_OPERATIONAL:recall\endcsname{40.0\%}
\expandafter\def\csname Num:2025:D30_AVAILABILITY_B05:EPSS_OPERATIONAL:yield\endcsname{0.83\%}
\expandafter\def\csname Num:2025:D30_AVAILABILITY_B10:COMMON:N\endcsname{38,368}
\expandafter\def\csname Num:2025:D30_AVAILABILITY_B10:COMMON:k\endcsname{3,837}
\expandafter\def\csname Num:2025:D30_AVAILABILITY_B10:COMMON:P\endcsname{40}
\expandafter\def\csname Num:2025:D30_AVAILABILITY_B10:CVSS_OPERATIONAL:captured\endcsname{9}
\expandafter\def\csname Num:2025:D30_AVAILABILITY_B10:CVSS_OPERATIONAL:minimum\endcsname{3}
\expandafter\def\csname Num:2025:D30_AVAILABILITY_B10:CVSS_OPERATIONAL:maximum\endcsname{9}
\expandafter\def\csname Num:2025:D30_AVAILABILITY_B10:CVSS_OPERATIONAL:recall\endcsname{22.5\%}
\expandafter\def\csname Num:2025:D30_AVAILABILITY_B10:CVSS_OPERATIONAL:yield\endcsname{0.23\%}
\expandafter\def\csname Num:2025:D30_AVAILABILITY_B10:EPSS_OPERATIONAL:captured\endcsname{20}
\expandafter\def\csname Num:2025:D30_AVAILABILITY_B10:EPSS_OPERATIONAL:minimum\endcsname{20}
\expandafter\def\csname Num:2025:D30_AVAILABILITY_B10:EPSS_OPERATIONAL:maximum\endcsname{20}
\expandafter\def\csname Num:2025:D30_AVAILABILITY_B10:EPSS_OPERATIONAL:recall\endcsname{50.0\%}
\expandafter\def\csname Num:2025:D30_AVAILABILITY_B10:EPSS_OPERATIONAL:yield\endcsname{0.52\%}
\expandafter\def\csname Num:2025:D30_COMPLETE_CASE_B05:COMMON:N\endcsname{31,618}
\expandafter\def\csname Num:2025:D30_COMPLETE_CASE_B05:COMMON:k\endcsname{1,581}
\expandafter\def\csname Num:2025:D30_COMPLETE_CASE_B05:COMMON:P\endcsname{34}
\expandafter\def\csname Num:2025:D30_COMPLETE_CASE_B05:CVSS_OPERATIONAL:captured\endcsname{5}
\expandafter\def\csname Num:2025:D30_COMPLETE_CASE_B05:CVSS_OPERATIONAL:minimum\endcsname{3}
\expandafter\def\csname Num:2025:D30_COMPLETE_CASE_B05:CVSS_OPERATIONAL:maximum\endcsname{9}
\expandafter\def\csname Num:2025:D30_COMPLETE_CASE_B05:CVSS_OPERATIONAL:recall\endcsname{14.7\%}
\expandafter\def\csname Num:2025:D30_COMPLETE_CASE_B05:CVSS_OPERATIONAL:yield\endcsname{0.32\%}
\expandafter\def\csname Num:2025:D30_COMPLETE_CASE_B05:EPSS_OPERATIONAL:captured\endcsname{12}
\expandafter\def\csname Num:2025:D30_COMPLETE_CASE_B05:EPSS_OPERATIONAL:minimum\endcsname{12}
\expandafter\def\csname Num:2025:D30_COMPLETE_CASE_B05:EPSS_OPERATIONAL:maximum\endcsname{12}
\expandafter\def\csname Num:2025:D30_COMPLETE_CASE_B05:EPSS_OPERATIONAL:recall\endcsname{35.3\%}
\expandafter\def\csname Num:2025:D30_COMPLETE_CASE_B05:EPSS_OPERATIONAL:yield\endcsname{0.76\%}
\expandafter\def\csname Num:2025:D30_NVD_SOURCE_B05:COMMON:N\endcsname{38,368}
\expandafter\def\csname Num:2025:D30_NVD_SOURCE_B05:COMMON:k\endcsname{1,919}
\expandafter\def\csname Num:2025:D30_NVD_SOURCE_B05:COMMON:P\endcsname{40}
\expandafter\def\csname Num:2025:D30_NVD_SOURCE_B05:CVSS_OPERATIONAL:captured\endcsname{1}
\expandafter\def\csname Num:2025:D30_NVD_SOURCE_B05:CVSS_OPERATIONAL:minimum\endcsname{0}
\expandafter\def\csname Num:2025:D30_NVD_SOURCE_B05:CVSS_OPERATIONAL:maximum\endcsname{1}
\expandafter\def\csname Num:2025:D30_NVD_SOURCE_B05:CVSS_OPERATIONAL:recall\endcsname{2.5\%}
\expandafter\def\csname Num:2025:D30_NVD_SOURCE_B05:CVSS_OPERATIONAL:yield\endcsname{0.05\%}
\expandafter\def\csname Num:2025:D30_NVD_SOURCE_B05:EPSS_OPERATIONAL:captured\endcsname{16}
\expandafter\def\csname Num:2025:D30_NVD_SOURCE_B05:EPSS_OPERATIONAL:minimum\endcsname{16}
\expandafter\def\csname Num:2025:D30_NVD_SOURCE_B05:EPSS_OPERATIONAL:maximum\endcsname{16}
\expandafter\def\csname Num:2025:D30_NVD_SOURCE_B05:EPSS_OPERATIONAL:recall\endcsname{40.0\%}
\expandafter\def\csname Num:2025:D30_NVD_SOURCE_B05:EPSS_OPERATIONAL:yield\endcsname{0.83\%}
\expandafter\def\csname Num:2025:D30_V31_ONLY_B05:COMMON:N\endcsname{38,368}
\expandafter\def\csname Num:2025:D30_V31_ONLY_B05:COMMON:k\endcsname{1,919}
\expandafter\def\csname Num:2025:D30_V31_ONLY_B05:COMMON:P\endcsname{40}
\expandafter\def\csname Num:2025:D30_V31_ONLY_B05:CVSS_OPERATIONAL:captured\endcsname{5}
\expandafter\def\csname Num:2025:D30_V31_ONLY_B05:CVSS_OPERATIONAL:minimum\endcsname{3}
\expandafter\def\csname Num:2025:D30_V31_ONLY_B05:CVSS_OPERATIONAL:maximum\endcsname{9}
\expandafter\def\csname Num:2025:D30_V31_ONLY_B05:CVSS_OPERATIONAL:recall\endcsname{12.5\%}
\expandafter\def\csname Num:2025:D30_V31_ONLY_B05:CVSS_OPERATIONAL:yield\endcsname{0.26\%}
\expandafter\def\csname Num:2025:D30_V31_ONLY_B05:EPSS_OPERATIONAL:captured\endcsname{16}
\expandafter\def\csname Num:2025:D30_V31_ONLY_B05:EPSS_OPERATIONAL:minimum\endcsname{16}
\expandafter\def\csname Num:2025:D30_V31_ONLY_B05:EPSS_OPERATIONAL:maximum\endcsname{16}
\expandafter\def\csname Num:2025:D30_V31_ONLY_B05:EPSS_OPERATIONAL:recall\endcsname{40.0\%}
\expandafter\def\csname Num:2025:D30_V31_ONLY_B05:EPSS_OPERATIONAL:yield\endcsname{0.83\%}
\expandafter\def\csname Num:2025:D7_AVAILABILITY_B05:COMMON:N\endcsname{38,476}
\expandafter\def\csname Num:2025:D7_AVAILABILITY_B05:COMMON:k\endcsname{1,924}
\expandafter\def\csname Num:2025:D7_AVAILABILITY_B05:COMMON:P\endcsname{60}
\expandafter\def\csname Num:2025:D7_AVAILABILITY_B05:CVSS_OPERATIONAL:captured\endcsname{14}
\expandafter\def\csname Num:2025:D7_AVAILABILITY_B05:CVSS_OPERATIONAL:minimum\endcsname{8}
\expandafter\def\csname Num:2025:D7_AVAILABILITY_B05:CVSS_OPERATIONAL:maximum\endcsname{21}
\expandafter\def\csname Num:2025:D7_AVAILABILITY_B05:CVSS_OPERATIONAL:recall\endcsname{23.3\%}
\expandafter\def\csname Num:2025:D7_AVAILABILITY_B05:CVSS_OPERATIONAL:yield\endcsname{0.73\%}
\expandafter\def\csname Num:2025:D7_AVAILABILITY_B05:EPSS_OPERATIONAL:captured\endcsname{23}
\expandafter\def\csname Num:2025:D7_AVAILABILITY_B05:EPSS_OPERATIONAL:minimum\endcsname{23}
\expandafter\def\csname Num:2025:D7_AVAILABILITY_B05:EPSS_OPERATIONAL:maximum\endcsname{24}
\expandafter\def\csname Num:2025:D7_AVAILABILITY_B05:EPSS_OPERATIONAL:recall\endcsname{38.3\%}
\expandafter\def\csname Num:2025:D7_AVAILABILITY_B05:EPSS_OPERATIONAL:yield\endcsname{1.20\%}
\expandafter\def\csname Num:Q1_2026:D30_AVAILABILITY_B01:COMMON:N\endcsname{15,141}
\expandafter\def\csname Num:Q1_2026:D30_AVAILABILITY_B01:COMMON:k\endcsname{152}
\expandafter\def\csname Num:Q1_2026:D30_AVAILABILITY_B01:COMMON:P\endcsname{11}
\expandafter\def\csname Num:Q1_2026:D30_AVAILABILITY_B01:CVSS_OPERATIONAL:captured\endcsname{1}
\expandafter\def\csname Num:Q1_2026:D30_AVAILABILITY_B01:CVSS_OPERATIONAL:minimum\endcsname{1}
\expandafter\def\csname Num:Q1_2026:D30_AVAILABILITY_B01:CVSS_OPERATIONAL:maximum\endcsname{1}
\expandafter\def\csname Num:Q1_2026:D30_AVAILABILITY_B01:CVSS_OPERATIONAL:recall\endcsname{9.1\%}
\expandafter\def\csname Num:Q1_2026:D30_AVAILABILITY_B01:CVSS_OPERATIONAL:yield\endcsname{0.66\%}
\expandafter\def\csname Num:Q1_2026:D30_AVAILABILITY_B01:EPSS_OPERATIONAL:captured\endcsname{2}
\expandafter\def\csname Num:Q1_2026:D30_AVAILABILITY_B01:EPSS_OPERATIONAL:minimum\endcsname{2}
\expandafter\def\csname Num:Q1_2026:D30_AVAILABILITY_B01:EPSS_OPERATIONAL:maximum\endcsname{2}
\expandafter\def\csname Num:Q1_2026:D30_AVAILABILITY_B01:EPSS_OPERATIONAL:recall\endcsname{18.2\%}
\expandafter\def\csname Num:Q1_2026:D30_AVAILABILITY_B01:EPSS_OPERATIONAL:yield\endcsname{1.32\%}
\expandafter\def\csname Num:Q1_2026:D30_AVAILABILITY_B05:COMMON:N\endcsname{15,141}
\expandafter\def\csname Num:Q1_2026:D30_AVAILABILITY_B05:COMMON:k\endcsname{758}
\expandafter\def\csname Num:Q1_2026:D30_AVAILABILITY_B05:COMMON:P\endcsname{11}
\expandafter\def\csname Num:Q1_2026:D30_AVAILABILITY_B05:COMMON:base_rate\endcsname{0.07\%}
\expandafter\def\csname Num:Q1_2026:D30_AVAILABILITY_B05:COMMON:neither\endcsname{6}
\expandafter\def\csname Num:Q1_2026:D30_AVAILABILITY_B05:CVSS_OPERATIONAL:captured\endcsname{4}
\expandafter\def\csname Num:Q1_2026:D30_AVAILABILITY_B05:CVSS_OPERATIONAL:minimum\endcsname{1}
\expandafter\def\csname Num:Q1_2026:D30_AVAILABILITY_B05:CVSS_OPERATIONAL:maximum\endcsname{5}
\expandafter\def\csname Num:Q1_2026:D30_AVAILABILITY_B05:CVSS_OPERATIONAL:recall\endcsname{36.4\%}
\expandafter\def\csname Num:Q1_2026:D30_AVAILABILITY_B05:CVSS_OPERATIONAL:yield\endcsname{0.53\%}
\expandafter\def\csname Num:Q1_2026:D30_AVAILABILITY_B05:EPSS_OPERATIONAL:captured\endcsname{3}
\expandafter\def\csname Num:Q1_2026:D30_AVAILABILITY_B05:EPSS_OPERATIONAL:minimum\endcsname{3}
\expandafter\def\csname Num:Q1_2026:D30_AVAILABILITY_B05:EPSS_OPERATIONAL:maximum\endcsname{3}
\expandafter\def\csname Num:Q1_2026:D30_AVAILABILITY_B05:EPSS_OPERATIONAL:recall\endcsname{27.3\%}
\expandafter\def\csname Num:Q1_2026:D30_AVAILABILITY_B05:EPSS_OPERATIONAL:yield\endcsname{0.40\%}
\expandafter\def\csname Num:Q1_2026:D30_AVAILABILITY_B10:COMMON:N\endcsname{15,141}
\expandafter\def\csname Num:Q1_2026:D30_AVAILABILITY_B10:COMMON:k\endcsname{1,515}
\expandafter\def\csname Num:Q1_2026:D30_AVAILABILITY_B10:COMMON:P\endcsname{11}
\expandafter\def\csname Num:Q1_2026:D30_AVAILABILITY_B10:CVSS_OPERATIONAL:captured\endcsname{5}
\expandafter\def\csname Num:Q1_2026:D30_AVAILABILITY_B10:CVSS_OPERATIONAL:minimum\endcsname{1}
\expandafter\def\csname Num:Q1_2026:D30_AVAILABILITY_B10:CVSS_OPERATIONAL:maximum\endcsname{5}
\expandafter\def\csname Num:Q1_2026:D30_AVAILABILITY_B10:CVSS_OPERATIONAL:recall\endcsname{45.5\%}
\expandafter\def\csname Num:Q1_2026:D30_AVAILABILITY_B10:CVSS_OPERATIONAL:yield\endcsname{0.33\%}
\expandafter\def\csname Num:Q1_2026:D30_AVAILABILITY_B10:EPSS_OPERATIONAL:captured\endcsname{3}
\expandafter\def\csname Num:Q1_2026:D30_AVAILABILITY_B10:EPSS_OPERATIONAL:minimum\endcsname{3}
\expandafter\def\csname Num:Q1_2026:D30_AVAILABILITY_B10:EPSS_OPERATIONAL:maximum\endcsname{3}
\expandafter\def\csname Num:Q1_2026:D30_AVAILABILITY_B10:EPSS_OPERATIONAL:recall\endcsname{27.3\%}
\expandafter\def\csname Num:Q1_2026:D30_AVAILABILITY_B10:EPSS_OPERATIONAL:yield\endcsname{0.20\%}
\expandafter\def\csname Num:Q1_2026:D30_COMPLETE_CASE_B05:COMMON:N\endcsname{13,934}
\expandafter\def\csname Num:Q1_2026:D30_COMPLETE_CASE_B05:COMMON:k\endcsname{697}
\expandafter\def\csname Num:Q1_2026:D30_COMPLETE_CASE_B05:COMMON:P\endcsname{11}
\expandafter\def\csname Num:Q1_2026:D30_COMPLETE_CASE_B05:CVSS_OPERATIONAL:captured\endcsname{4}
\expandafter\def\csname Num:Q1_2026:D30_COMPLETE_CASE_B05:CVSS_OPERATIONAL:minimum\endcsname{1}
\expandafter\def\csname Num:Q1_2026:D30_COMPLETE_CASE_B05:CVSS_OPERATIONAL:maximum\endcsname{5}
\expandafter\def\csname Num:Q1_2026:D30_COMPLETE_CASE_B05:CVSS_OPERATIONAL:recall\endcsname{36.4\%}
\expandafter\def\csname Num:Q1_2026:D30_COMPLETE_CASE_B05:CVSS_OPERATIONAL:yield\endcsname{0.57\%}
\expandafter\def\csname Num:Q1_2026:D30_COMPLETE_CASE_B05:EPSS_OPERATIONAL:captured\endcsname{3}
\expandafter\def\csname Num:Q1_2026:D30_COMPLETE_CASE_B05:EPSS_OPERATIONAL:minimum\endcsname{3}
\expandafter\def\csname Num:Q1_2026:D30_COMPLETE_CASE_B05:EPSS_OPERATIONAL:maximum\endcsname{3}
\expandafter\def\csname Num:Q1_2026:D30_COMPLETE_CASE_B05:EPSS_OPERATIONAL:recall\endcsname{27.3\%}
\expandafter\def\csname Num:Q1_2026:D30_COMPLETE_CASE_B05:EPSS_OPERATIONAL:yield\endcsname{0.43\%}
\expandafter\def\csname Num:Q1_2026:D30_NVD_SOURCE_B05:COMMON:N\endcsname{15,141}
\expandafter\def\csname Num:Q1_2026:D30_NVD_SOURCE_B05:COMMON:k\endcsname{758}
\expandafter\def\csname Num:Q1_2026:D30_NVD_SOURCE_B05:COMMON:P\endcsname{11}
\expandafter\def\csname Num:Q1_2026:D30_NVD_SOURCE_B05:CVSS_OPERATIONAL:captured\endcsname{0}
\expandafter\def\csname Num:Q1_2026:D30_NVD_SOURCE_B05:CVSS_OPERATIONAL:minimum\endcsname{0}
\expandafter\def\csname Num:Q1_2026:D30_NVD_SOURCE_B05:CVSS_OPERATIONAL:maximum\endcsname{0}
\expandafter\def\csname Num:Q1_2026:D30_NVD_SOURCE_B05:CVSS_OPERATIONAL:recall\endcsname{0.0\%}
\expandafter\def\csname Num:Q1_2026:D30_NVD_SOURCE_B05:CVSS_OPERATIONAL:yield\endcsname{0.00\%}
\expandafter\def\csname Num:Q1_2026:D30_NVD_SOURCE_B05:EPSS_OPERATIONAL:captured\endcsname{3}
\expandafter\def\csname Num:Q1_2026:D30_NVD_SOURCE_B05:EPSS_OPERATIONAL:minimum\endcsname{3}
\expandafter\def\csname Num:Q1_2026:D30_NVD_SOURCE_B05:EPSS_OPERATIONAL:maximum\endcsname{3}
\expandafter\def\csname Num:Q1_2026:D30_NVD_SOURCE_B05:EPSS_OPERATIONAL:recall\endcsname{27.3\%}
\expandafter\def\csname Num:Q1_2026:D30_NVD_SOURCE_B05:EPSS_OPERATIONAL:yield\endcsname{0.40\%}
\expandafter\def\csname Num:Q1_2026:D30_V31_ONLY_B05:COMMON:N\endcsname{15,141}
\expandafter\def\csname Num:Q1_2026:D30_V31_ONLY_B05:COMMON:k\endcsname{758}
\expandafter\def\csname Num:Q1_2026:D30_V31_ONLY_B05:COMMON:P\endcsname{11}
\expandafter\def\csname Num:Q1_2026:D30_V31_ONLY_B05:CVSS_OPERATIONAL:captured\endcsname{3}
\expandafter\def\csname Num:Q1_2026:D30_V31_ONLY_B05:CVSS_OPERATIONAL:minimum\endcsname{1}
\expandafter\def\csname Num:Q1_2026:D30_V31_ONLY_B05:CVSS_OPERATIONAL:maximum\endcsname{4}
\expandafter\def\csname Num:Q1_2026:D30_V31_ONLY_B05:CVSS_OPERATIONAL:recall\endcsname{27.3\%}
\expandafter\def\csname Num:Q1_2026:D30_V31_ONLY_B05:CVSS_OPERATIONAL:yield\endcsname{0.40\%}
\expandafter\def\csname Num:Q1_2026:D30_V31_ONLY_B05:EPSS_OPERATIONAL:captured\endcsname{3}
\expandafter\def\csname Num:Q1_2026:D30_V31_ONLY_B05:EPSS_OPERATIONAL:minimum\endcsname{3}
\expandafter\def\csname Num:Q1_2026:D30_V31_ONLY_B05:EPSS_OPERATIONAL:maximum\endcsname{3}
\expandafter\def\csname Num:Q1_2026:D30_V31_ONLY_B05:EPSS_OPERATIONAL:recall\endcsname{27.3\%}
\expandafter\def\csname Num:Q1_2026:D30_V31_ONLY_B05:EPSS_OPERATIONAL:yield\endcsname{0.40\%}
\expandafter\def\csname Num:Q1_2026:D7_AVAILABILITY_B05:COMMON:N\endcsname{15,160}
\expandafter\def\csname Num:Q1_2026:D7_AVAILABILITY_B05:COMMON:k\endcsname{758}
\expandafter\def\csname Num:Q1_2026:D7_AVAILABILITY_B05:COMMON:P\endcsname{16}
\expandafter\def\csname Num:Q1_2026:D7_AVAILABILITY_B05:CVSS_OPERATIONAL:captured\endcsname{5}
\expandafter\def\csname Num:Q1_2026:D7_AVAILABILITY_B05:CVSS_OPERATIONAL:minimum\endcsname{1}
\expandafter\def\csname Num:Q1_2026:D7_AVAILABILITY_B05:CVSS_OPERATIONAL:maximum\endcsname{6}
\expandafter\def\csname Num:Q1_2026:D7_AVAILABILITY_B05:CVSS_OPERATIONAL:recall\endcsname{31.3\%}
\expandafter\def\csname Num:Q1_2026:D7_AVAILABILITY_B05:CVSS_OPERATIONAL:yield\endcsname{0.66\%}
\expandafter\def\csname Num:Q1_2026:D7_AVAILABILITY_B05:EPSS_OPERATIONAL:captured\endcsname{6}
\expandafter\def\csname Num:Q1_2026:D7_AVAILABILITY_B05:EPSS_OPERATIONAL:minimum\endcsname{6}
\expandafter\def\csname Num:Q1_2026:D7_AVAILABILITY_B05:EPSS_OPERATIONAL:maximum\endcsname{6}
\expandafter\def\csname Num:Q1_2026:D7_AVAILABILITY_B05:EPSS_OPERATIONAL:recall\endcsname{37.5\%}
\expandafter\def\csname Num:Q1_2026:D7_AVAILABILITY_B05:EPSS_OPERATIONAL:yield\endcsname{0.79\%}
\expandafter\def\csname Num:2025:D30_AVAILABILITY_B05:CVSS_OPERATIONAL:expected\endcsname{5.7}
\expandafter\def\csname Num:2025:D30_AVAILABILITY_B05:EPSS_OPERATIONAL:expected\endcsname{16.0}
\expandafter\def\csname Num:Q1_2026:D30_AVAILABILITY_B05:CVSS_OPERATIONAL:expected\endcsname{2.6}
\expandafter\def\csname Num:Q1_2026:D30_AVAILABILITY_B05:EPSS_OPERATIONAL:expected\endcsname{3.0}
\expandafter\def\csname Diag:2025:positives_D7\endcsname{60}
\expandafter\def\csname Diag:2025:positives_D30\endcsname{40}
\expandafter\def\csname Diag:2025:common\endcsname{38}
\expandafter\def\csname Diag:2025:D7_only\endcsname{22}
\expandafter\def\csname Diag:2025:D30_only\endcsname{2}
\expandafter\def\csname Diag:2025:union_n\endcsname{62}
\expandafter\def\csname Diag:2025:early_subgroup_n\endcsname{22}
\expandafter\def\csname Diag:Q1_2026:positives_D7\endcsname{16}
\expandafter\def\csname Diag:Q1_2026:positives_D30\endcsname{11}
\expandafter\def\csname Diag:Q1_2026:common\endcsname{10}
\expandafter\def\csname Diag:Q1_2026:D7_only\endcsname{6}
\expandafter\def\csname Diag:Q1_2026:D30_only\endcsname{1}
\expandafter\def\csname Diag:Q1_2026:union_n\endcsname{17}
\expandafter\def\csname Diag:Q1_2026:early_subgroup_n\endcsname{6}
\expandafter\def\csname Data:2025:D7:scope_n\endcsname{39,917}
\expandafter\def\csname Data:2025:D7:rejected\endcsname{1,371}
\expandafter\def\csname Data:2025:D7:known_kev\endcsname{70}
\expandafter\def\csname Data:2025:D7:other\endcsname{0}
\expandafter\def\csname Data:2025:D7:N\endcsname{38,476}
\expandafter\def\csname Data:2025:D7:P\endcsname{60}
\expandafter\def\csname Data:2025:D7:complete_cases\endcsname{30,741}
\expandafter\def\csname Data:2025:D7:cvss_only\endcsname{0}
\expandafter\def\csname Data:2025:D7:epss_only\endcsname{7,735}
\expandafter\def\csname Data:2025:D30:scope_n\endcsname{39,917}
\expandafter\def\csname Data:2025:D30:rejected\endcsname{1,457}
\expandafter\def\csname Data:2025:D30:known_kev\endcsname{92}
\expandafter\def\csname Data:2025:D30:other\endcsname{0}
\expandafter\def\csname Data:2025:D30:N\endcsname{38,368}
\expandafter\def\csname Data:2025:D30:P\endcsname{40}
\expandafter\def\csname Data:2025:D30:complete_cases\endcsname{31,618}
\expandafter\def\csname Data:2025:D30:cvss_only\endcsname{0}
\expandafter\def\csname Data:2025:D30:epss_only\endcsname{6,750}
\expandafter\def\csname Data:Q1_2026:D7:scope_n\endcsname{16,255}
\expandafter\def\csname Data:Q1_2026:D7:rejected\endcsname{1,068}
\expandafter\def\csname Data:Q1_2026:D7:known_kev\endcsname{27}
\expandafter\def\csname Data:Q1_2026:D7:other\endcsname{0}
\expandafter\def\csname Data:Q1_2026:D7:N\endcsname{15,160}
\expandafter\def\csname Data:Q1_2026:D7:P\endcsname{16}
\expandafter\def\csname Data:Q1_2026:D7:complete_cases\endcsname{13,258}
\expandafter\def\csname Data:Q1_2026:D7:cvss_only\endcsname{0}
\expandafter\def\csname Data:Q1_2026:D7:epss_only\endcsname{1,902}
\expandafter\def\csname Data:Q1_2026:D30:scope_n\endcsname{16,255}
\expandafter\def\csname Data:Q1_2026:D30:rejected\endcsname{1,081}
\expandafter\def\csname Data:Q1_2026:D30:known_kev\endcsname{33}
\expandafter\def\csname Data:Q1_2026:D30:other\endcsname{0}
\expandafter\def\csname Data:Q1_2026:D30:N\endcsname{15,141}
\expandafter\def\csname Data:Q1_2026:D30:P\endcsname{11}
\expandafter\def\csname Data:Q1_2026:D30:complete_cases\endcsname{13,934}
\expandafter\def\csname Data:Q1_2026:D30:cvss_only\endcsname{0}
\expandafter\def\csname Data:Q1_2026:D30:epss_only\endcsname{1,207}
\expandafter\def\csname Provenance:2025:NVD_PRIMARY:count\endcsname{7,507}
\expandafter\def\csname Provenance:2025:NVD_PRIMARY:share\endcsname{19.6\%}
\expandafter\def\csname Provenance:2025:CNA:count\endcsname{24,111}
\expandafter\def\csname Provenance:2025:CNA:share\endcsname{62.8\%}
\expandafter\def\csname Provenance:2025:NONE:count\endcsname{6,750}
\expandafter\def\csname Provenance:2025:NONE:share\endcsname{17.6\%}
\expandafter\def\csname NVD:2025:positive:available\endcsname{6}
\expandafter\def\csname NVD:2025:positive:absent\endcsname{34}
\expandafter\def\csname NVD:2025:positive:unresolved\endcsname{0}
\expandafter\def\csname NVD:2025:positive:denominator\endcsname{40}
\expandafter\def\csname NVD:2025:positive:share\endcsname{15.0\%}
\expandafter\def\csname NVD:2025:negative:available\endcsname{7,501}
\expandafter\def\csname NVD:2025:negative:absent\endcsname{30,827}
\expandafter\def\csname NVD:2025:negative:unresolved\endcsname{0}
\expandafter\def\csname NVD:2025:negative:denominator\endcsname{38,328}
\expandafter\def\csname NVD:2025:negative:share\endcsname{19.6\%}
\expandafter\def\csname Stability:2025:CVSS_OPERATIONAL:N\endcsname{38,362}
\expandafter\def\csname Stability:2025:CVSS_OPERATIONAL:k\endcsname{1,919}
\expandafter\def\csname Stability:2025:CVSS_OPERATIONAL:stayed\endcsname{1,447}
\expandafter\def\csname Stability:2025:CVSS_OPERATIONAL:entered\endcsname{472}
\expandafter\def\csname Stability:2025:CVSS_OPERATIONAL:exited\endcsname{472}
\expandafter\def\csname Stability:2025:CVSS_OPERATIONAL:turnover\endcsname{24.6\%}
\expandafter\def\csname Stability:2025:CVSS_OPERATIONAL:median\endcsname{0.028}
\expandafter\def\csname Stability:2025:CVSS_OPERATIONAL:p90\endcsname{0.036}
\expandafter\def\csname Stability:2025:EPSS_OPERATIONAL:N\endcsname{38,362}
\expandafter\def\csname Stability:2025:EPSS_OPERATIONAL:k\endcsname{1,919}
\expandafter\def\csname Stability:2025:EPSS_OPERATIONAL:stayed\endcsname{1,659}
\expandafter\def\csname Stability:2025:EPSS_OPERATIONAL:entered\endcsname{260}
\expandafter\def\csname Stability:2025:EPSS_OPERATIONAL:exited\endcsname{260}
\expandafter\def\csname Stability:2025:EPSS_OPERATIONAL:turnover\endcsname{13.5\%}
\expandafter\def\csname Stability:2025:EPSS_OPERATIONAL:median\endcsname{0.005}
\expandafter\def\csname Stability:2025:EPSS_OPERATIONAL:p90\endcsname{0.076}
\expandafter\def\csname Provenance:Q1_2026:NVD_PRIMARY:count\endcsname{4,897}
\expandafter\def\csname Provenance:Q1_2026:NVD_PRIMARY:share\endcsname{32.3\%}
\expandafter\def\csname Provenance:Q1_2026:CNA:count\endcsname{9,037}
\expandafter\def\csname Provenance:Q1_2026:CNA:share\endcsname{59.7\%}
\expandafter\def\csname Provenance:Q1_2026:NONE:count\endcsname{1,207}
\expandafter\def\csname Provenance:Q1_2026:NONE:share\endcsname{8.0\%}
\expandafter\def\csname NVD:Q1_2026:positive:available\endcsname{2}
\expandafter\def\csname NVD:Q1_2026:positive:absent\endcsname{9}
\expandafter\def\csname NVD:Q1_2026:positive:unresolved\endcsname{0}
\expandafter\def\csname NVD:Q1_2026:positive:denominator\endcsname{11}
\expandafter\def\csname NVD:Q1_2026:positive:share\endcsname{18.2\%}
\expandafter\def\csname NVD:Q1_2026:negative:available\endcsname{4,895}
\expandafter\def\csname NVD:Q1_2026:negative:absent\endcsname{10,235}
\expandafter\def\csname NVD:Q1_2026:negative:unresolved\endcsname{0}
\expandafter\def\csname NVD:Q1_2026:negative:denominator\endcsname{15,130}
\expandafter\def\csname NVD:Q1_2026:negative:share\endcsname{32.4\%}
\expandafter\def\csname Stability:Q1_2026:CVSS_OPERATIONAL:N\endcsname{15,141}
\expandafter\def\csname Stability:Q1_2026:CVSS_OPERATIONAL:k\endcsname{758}
\expandafter\def\csname Stability:Q1_2026:CVSS_OPERATIONAL:stayed\endcsname{606}
\expandafter\def\csname Stability:Q1_2026:CVSS_OPERATIONAL:entered\endcsname{152}
\expandafter\def\csname Stability:Q1_2026:CVSS_OPERATIONAL:exited\endcsname{152}
\expandafter\def\csname Stability:Q1_2026:CVSS_OPERATIONAL:turnover\endcsname{20.1\%}
\expandafter\def\csname Stability:Q1_2026:CVSS_OPERATIONAL:median\endcsname{0.034}
\expandafter\def\csname Stability:Q1_2026:CVSS_OPERATIONAL:p90\endcsname{0.051}
\expandafter\def\csname Stability:Q1_2026:EPSS_OPERATIONAL:N\endcsname{15,141}
\expandafter\def\csname Stability:Q1_2026:EPSS_OPERATIONAL:k\endcsname{758}
\expandafter\def\csname Stability:Q1_2026:EPSS_OPERATIONAL:stayed\endcsname{641}
\expandafter\def\csname Stability:Q1_2026:EPSS_OPERATIONAL:entered\endcsname{117}
\expandafter\def\csname Stability:Q1_2026:EPSS_OPERATIONAL:exited\endcsname{117}
\expandafter\def\csname Stability:Q1_2026:EPSS_OPERATIONAL:turnover\endcsname{15.4\%}
\expandafter\def\csname Stability:Q1_2026:EPSS_OPERATIONAL:median\endcsname{0.039}
\expandafter\def\csname Stability:Q1_2026:EPSS_OPERATIONAL:p90\endcsname{0.187}
\expandafter\def\csname Block:Q1_2026:1:N\endcsname{6,853}
\expandafter\def\csname Block:Q1_2026:1:P\endcsname{6}
\expandafter\def\csname Block:Q1_2026:1:selected_cvss\endcsname{666}
\expandafter\def\csname Block:Q1_2026:1:selected_epss\endcsname{349}
\expandafter\def\csname Block:Q1_2026:1:captured_cvss\endcsname{3}
\expandafter\def\csname Block:Q1_2026:1:captured_epss\endcsname{3}
\expandafter\def\csname Block:Q1_2026:1:both\endcsname{2}
\expandafter\def\csname Block:Q1_2026:1:cvss_only\endcsname{1}
\expandafter\def\csname Block:Q1_2026:1:epss_only\endcsname{1}
\expandafter\def\csname Block:Q1_2026:1:neither\endcsname{2}
\expandafter\def\csname Block:Q1_2026:1:start_date\endcsname{1 January 2026}
\expandafter\def\csname Block:Q1_2026:1:end_date\endcsname{17 February 2026}
\expandafter\def\csname Block:Q1_2026:2:N\endcsname{8,288}
\expandafter\def\csname Block:Q1_2026:2:P\endcsname{5}
\expandafter\def\csname Block:Q1_2026:2:selected_cvss\endcsname{92}
\expandafter\def\csname Block:Q1_2026:2:selected_epss\endcsname{409}
\expandafter\def\csname Block:Q1_2026:2:captured_cvss\endcsname{1}
\expandafter\def\csname Block:Q1_2026:2:captured_epss\endcsname{0}
\expandafter\def\csname Block:Q1_2026:2:both\endcsname{0}
\expandafter\def\csname Block:Q1_2026:2:cvss_only\endcsname{1}
\expandafter\def\csname Block:Q1_2026:2:epss_only\endcsname{0}
\expandafter\def\csname Block:Q1_2026:2:neither\endcsname{4}
\expandafter\def\csname Block:Q1_2026:2:start_date\endcsname{18 February 2026}
\expandafter\def\csname Block:Q1_2026:2:end_date\endcsname{31 March 2026}
\expandafter\def\csname Early:2025:subgroup_n\endcsname{22}
\expandafter\def\csname Early:2025:both\endcsname{7}
\expandafter\def\csname Early:2025:cvss_only\endcsname{3}
\expandafter\def\csname Early:2025:epss_only\endcsname{3}
\expandafter\def\csname Early:2025:neither\endcsname{9}
\expandafter\def\csname Early:2025:captured_cvss\endcsname{10}
\expandafter\def\csname Early:2025:captured_epss\endcsname{10}
\expandafter\def\csname Early:Q1_2026:subgroup_n\endcsname{6}
\expandafter\def\csname Early:Q1_2026:both\endcsname{2}
\expandafter\def\csname Early:Q1_2026:cvss_only\endcsname{0}
\expandafter\def\csname Early:Q1_2026:epss_only\endcsname{2}
\expandafter\def\csname Early:Q1_2026:neither\endcsname{2}
\expandafter\def\csname Early:Q1_2026:captured_cvss\endcsname{2}
\expandafter\def\csname Early:Q1_2026:captured_epss\endcsname{4}
\expandafter\def\csname Tie:2025:D30_AVAILABILITY_B01:CVSS_OPERATIONAL:cutoff_score\endcsname{9.8}
\expandafter\def\csname Tie:2025:D30_AVAILABILITY_B01:CVSS_OPERATIONAL:A\endcsname{241}
\expandafter\def\csname Tie:2025:D30_AVAILABILITY_B01:CVSS_OPERATIONAL:T\endcsname{3,743}
\expandafter\def\csname Tie:2025:D30_AVAILABILITY_B01:CVSS_OPERATIONAL:r\endcsname{143}
\expandafter\def\csname Tie:2025:D30_AVAILABILITY_B01:CVSS_OPERATIONAL:a\endcsname{3}
\expandafter\def\csname Tie:2025:D30_AVAILABILITY_B01:CVSS_OPERATIONAL:q\endcsname{6}
\expandafter\def\csname Tie:2025:D30_AVAILABILITY_B01:CVSS_OPERATIONAL:k\endcsname{384}
\expandafter\def\csname Tie:2025:D30_AVAILABILITY_B01:CVSS_OPERATIONAL:minimum\endcsname{3}
\expandafter\def\csname Tie:2025:D30_AVAILABILITY_B01:CVSS_OPERATIONAL:maximum\endcsname{9}
\expandafter\def\csname Tie:2025:D30_AVAILABILITY_B01:EPSS_OPERATIONAL:cutoff_score\endcsname{0.01711}
\expandafter\def\csname Tie:2025:D30_AVAILABILITY_B01:EPSS_OPERATIONAL:A\endcsname{383}
\expandafter\def\csname Tie:2025:D30_AVAILABILITY_B01:EPSS_OPERATIONAL:T\endcsname{1}
\expandafter\def\csname Tie:2025:D30_AVAILABILITY_B01:EPSS_OPERATIONAL:r\endcsname{1}
\expandafter\def\csname Tie:2025:D30_AVAILABILITY_B01:EPSS_OPERATIONAL:a\endcsname{8}
\expandafter\def\csname Tie:2025:D30_AVAILABILITY_B01:EPSS_OPERATIONAL:q\endcsname{0}
\expandafter\def\csname Tie:2025:D30_AVAILABILITY_B01:EPSS_OPERATIONAL:k\endcsname{384}
\expandafter\def\csname Tie:2025:D30_AVAILABILITY_B01:EPSS_OPERATIONAL:minimum\endcsname{8}
\expandafter\def\csname Tie:2025:D30_AVAILABILITY_B01:EPSS_OPERATIONAL:maximum\endcsname{8}
\expandafter\def\csname Tie:2025:D30_AVAILABILITY_B05:CVSS_OPERATIONAL:cutoff_score\endcsname{9.8}
\expandafter\def\csname Tie:2025:D30_AVAILABILITY_B05:CVSS_OPERATIONAL:A\endcsname{241}
\expandafter\def\csname Tie:2025:D30_AVAILABILITY_B05:CVSS_OPERATIONAL:T\endcsname{3,743}
\expandafter\def\csname Tie:2025:D30_AVAILABILITY_B05:CVSS_OPERATIONAL:r\endcsname{1,678}
\expandafter\def\csname Tie:2025:D30_AVAILABILITY_B05:CVSS_OPERATIONAL:a\endcsname{3}
\expandafter\def\csname Tie:2025:D30_AVAILABILITY_B05:CVSS_OPERATIONAL:q\endcsname{6}
\expandafter\def\csname Tie:2025:D30_AVAILABILITY_B05:CVSS_OPERATIONAL:k\endcsname{1,919}
\expandafter\def\csname Tie:2025:D30_AVAILABILITY_B05:CVSS_OPERATIONAL:minimum\endcsname{3}
\expandafter\def\csname Tie:2025:D30_AVAILABILITY_B05:CVSS_OPERATIONAL:maximum\endcsname{9}
\expandafter\def\csname Tie:2025:D30_AVAILABILITY_B05:EPSS_OPERATIONAL:cutoff_score\endcsname{0.00284}
\expandafter\def\csname Tie:2025:D30_AVAILABILITY_B05:EPSS_OPERATIONAL:A\endcsname{1,916}
\expandafter\def\csname Tie:2025:D30_AVAILABILITY_B05:EPSS_OPERATIONAL:T\endcsname{7}
\expandafter\def\csname Tie:2025:D30_AVAILABILITY_B05:EPSS_OPERATIONAL:r\endcsname{3}
\expandafter\def\csname Tie:2025:D30_AVAILABILITY_B05:EPSS_OPERATIONAL:a\endcsname{16}
\expandafter\def\csname Tie:2025:D30_AVAILABILITY_B05:EPSS_OPERATIONAL:q\endcsname{0}
\expandafter\def\csname Tie:2025:D30_AVAILABILITY_B05:EPSS_OPERATIONAL:k\endcsname{1,919}
\expandafter\def\csname Tie:2025:D30_AVAILABILITY_B05:EPSS_OPERATIONAL:minimum\endcsname{16}
\expandafter\def\csname Tie:2025:D30_AVAILABILITY_B05:EPSS_OPERATIONAL:maximum\endcsname{16}
\expandafter\def\csname Tie:2025:D30_AVAILABILITY_B10:CVSS_OPERATIONAL:cutoff_score\endcsname{9.8}
\expandafter\def\csname Tie:2025:D30_AVAILABILITY_B10:CVSS_OPERATIONAL:A\endcsname{241}
\expandafter\def\csname Tie:2025:D30_AVAILABILITY_B10:CVSS_OPERATIONAL:T\endcsname{3,743}
\expandafter\def\csname Tie:2025:D30_AVAILABILITY_B10:CVSS_OPERATIONAL:r\endcsname{3,596}
\expandafter\def\csname Tie:2025:D30_AVAILABILITY_B10:CVSS_OPERATIONAL:a\endcsname{3}
\expandafter\def\csname Tie:2025:D30_AVAILABILITY_B10:CVSS_OPERATIONAL:q\endcsname{6}
\expandafter\def\csname Tie:2025:D30_AVAILABILITY_B10:CVSS_OPERATIONAL:k\endcsname{3,837}
\expandafter\def\csname Tie:2025:D30_AVAILABILITY_B10:CVSS_OPERATIONAL:minimum\endcsname{3}
\expandafter\def\csname Tie:2025:D30_AVAILABILITY_B10:CVSS_OPERATIONAL:maximum\endcsname{9}
\expandafter\def\csname Tie:2025:D30_AVAILABILITY_B10:EPSS_OPERATIONAL:cutoff_score\endcsname{0.00132}
\expandafter\def\csname Tie:2025:D30_AVAILABILITY_B10:EPSS_OPERATIONAL:A\endcsname{3,832}
\expandafter\def\csname Tie:2025:D30_AVAILABILITY_B10:EPSS_OPERATIONAL:T\endcsname{19}
\expandafter\def\csname Tie:2025:D30_AVAILABILITY_B10:EPSS_OPERATIONAL:r\endcsname{5}
\expandafter\def\csname Tie:2025:D30_AVAILABILITY_B10:EPSS_OPERATIONAL:a\endcsname{20}
\expandafter\def\csname Tie:2025:D30_AVAILABILITY_B10:EPSS_OPERATIONAL:q\endcsname{0}
\expandafter\def\csname Tie:2025:D30_AVAILABILITY_B10:EPSS_OPERATIONAL:k\endcsname{3,837}
\expandafter\def\csname Tie:2025:D30_AVAILABILITY_B10:EPSS_OPERATIONAL:minimum\endcsname{20}
\expandafter\def\csname Tie:2025:D30_AVAILABILITY_B10:EPSS_OPERATIONAL:maximum\endcsname{20}
\expandafter\def\csname Tie:Q1_2026:D30_AVAILABILITY_B01:CVSS_OPERATIONAL:cutoff_score\endcsname{9.9}
\expandafter\def\csname Tie:Q1_2026:D30_AVAILABILITY_B01:CVSS_OPERATIONAL:A\endcsname{74}
\expandafter\def\csname Tie:Q1_2026:D30_AVAILABILITY_B01:CVSS_OPERATIONAL:T\endcsname{95}
\expandafter\def\csname Tie:Q1_2026:D30_AVAILABILITY_B01:CVSS_OPERATIONAL:r\endcsname{78}
\expandafter\def\csname Tie:Q1_2026:D30_AVAILABILITY_B01:CVSS_OPERATIONAL:a\endcsname{1}
\expandafter\def\csname Tie:Q1_2026:D30_AVAILABILITY_B01:CVSS_OPERATIONAL:q\endcsname{0}
\expandafter\def\csname Tie:Q1_2026:D30_AVAILABILITY_B01:CVSS_OPERATIONAL:k\endcsname{152}
\expandafter\def\csname Tie:Q1_2026:D30_AVAILABILITY_B01:CVSS_OPERATIONAL:minimum\endcsname{1}
\expandafter\def\csname Tie:Q1_2026:D30_AVAILABILITY_B01:CVSS_OPERATIONAL:maximum\endcsname{1}
\expandafter\def\csname Tie:Q1_2026:D30_AVAILABILITY_B01:EPSS_OPERATIONAL:cutoff_score\endcsname{0.01005}
\expandafter\def\csname Tie:Q1_2026:D30_AVAILABILITY_B01:EPSS_OPERATIONAL:A\endcsname{151}
\expandafter\def\csname Tie:Q1_2026:D30_AVAILABILITY_B01:EPSS_OPERATIONAL:T\endcsname{3}
\expandafter\def\csname Tie:Q1_2026:D30_AVAILABILITY_B01:EPSS_OPERATIONAL:r\endcsname{1}
\expandafter\def\csname Tie:Q1_2026:D30_AVAILABILITY_B01:EPSS_OPERATIONAL:a\endcsname{2}
\expandafter\def\csname Tie:Q1_2026:D30_AVAILABILITY_B01:EPSS_OPERATIONAL:q\endcsname{0}
\expandafter\def\csname Tie:Q1_2026:D30_AVAILABILITY_B01:EPSS_OPERATIONAL:k\endcsname{152}
\expandafter\def\csname Tie:Q1_2026:D30_AVAILABILITY_B01:EPSS_OPERATIONAL:minimum\endcsname{2}
\expandafter\def\csname Tie:Q1_2026:D30_AVAILABILITY_B01:EPSS_OPERATIONAL:maximum\endcsname{2}
\expandafter\def\csname Tie:Q1_2026:D30_AVAILABILITY_B05:CVSS_OPERATIONAL:cutoff_score\endcsname{9.8}
\expandafter\def\csname Tie:Q1_2026:D30_AVAILABILITY_B05:CVSS_OPERATIONAL:A\endcsname{169}
\expandafter\def\csname Tie:Q1_2026:D30_AVAILABILITY_B05:CVSS_OPERATIONAL:T\endcsname{1,434}
\expandafter\def\csname Tie:Q1_2026:D30_AVAILABILITY_B05:CVSS_OPERATIONAL:r\endcsname{589}
\expandafter\def\csname Tie:Q1_2026:D30_AVAILABILITY_B05:CVSS_OPERATIONAL:a\endcsname{1}
\expandafter\def\csname Tie:Q1_2026:D30_AVAILABILITY_B05:CVSS_OPERATIONAL:q\endcsname{4}
\expandafter\def\csname Tie:Q1_2026:D30_AVAILABILITY_B05:CVSS_OPERATIONAL:k\endcsname{758}
\expandafter\def\csname Tie:Q1_2026:D30_AVAILABILITY_B05:CVSS_OPERATIONAL:minimum\endcsname{1}
\expandafter\def\csname Tie:Q1_2026:D30_AVAILABILITY_B05:CVSS_OPERATIONAL:maximum\endcsname{5}
\expandafter\def\csname Tie:Q1_2026:D30_AVAILABILITY_B05:EPSS_OPERATIONAL:cutoff_score\endcsname{0.00240}
\expandafter\def\csname Tie:Q1_2026:D30_AVAILABILITY_B05:EPSS_OPERATIONAL:A\endcsname{754}
\expandafter\def\csname Tie:Q1_2026:D30_AVAILABILITY_B05:EPSS_OPERATIONAL:T\endcsname{5}
\expandafter\def\csname Tie:Q1_2026:D30_AVAILABILITY_B05:EPSS_OPERATIONAL:r\endcsname{4}
\expandafter\def\csname Tie:Q1_2026:D30_AVAILABILITY_B05:EPSS_OPERATIONAL:a\endcsname{3}
\expandafter\def\csname Tie:Q1_2026:D30_AVAILABILITY_B05:EPSS_OPERATIONAL:q\endcsname{0}
\expandafter\def\csname Tie:Q1_2026:D30_AVAILABILITY_B05:EPSS_OPERATIONAL:k\endcsname{758}
\expandafter\def\csname Tie:Q1_2026:D30_AVAILABILITY_B05:EPSS_OPERATIONAL:minimum\endcsname{3}
\expandafter\def\csname Tie:Q1_2026:D30_AVAILABILITY_B05:EPSS_OPERATIONAL:maximum\endcsname{3}
\expandafter\def\csname Tie:Q1_2026:D30_AVAILABILITY_B10:CVSS_OPERATIONAL:cutoff_score\endcsname{9.8}
\expandafter\def\csname Tie:Q1_2026:D30_AVAILABILITY_B10:CVSS_OPERATIONAL:A\endcsname{169}
\expandafter\def\csname Tie:Q1_2026:D30_AVAILABILITY_B10:CVSS_OPERATIONAL:T\endcsname{1,434}
\expandafter\def\csname Tie:Q1_2026:D30_AVAILABILITY_B10:CVSS_OPERATIONAL:r\endcsname{1,346}
\expandafter\def\csname Tie:Q1_2026:D30_AVAILABILITY_B10:CVSS_OPERATIONAL:a\endcsname{1}
\expandafter\def\csname Tie:Q1_2026:D30_AVAILABILITY_B10:CVSS_OPERATIONAL:q\endcsname{4}
\expandafter\def\csname Tie:Q1_2026:D30_AVAILABILITY_B10:CVSS_OPERATIONAL:k\endcsname{1,515}
\expandafter\def\csname Tie:Q1_2026:D30_AVAILABILITY_B10:CVSS_OPERATIONAL:minimum\endcsname{1}
\expandafter\def\csname Tie:Q1_2026:D30_AVAILABILITY_B10:CVSS_OPERATIONAL:maximum\endcsname{5}
\expandafter\def\csname Tie:Q1_2026:D30_AVAILABILITY_B10:EPSS_OPERATIONAL:cutoff_score\endcsname{0.00142}
\expandafter\def\csname Tie:Q1_2026:D30_AVAILABILITY_B10:EPSS_OPERATIONAL:A\endcsname{1,508}
\expandafter\def\csname Tie:Q1_2026:D30_AVAILABILITY_B10:EPSS_OPERATIONAL:T\endcsname{15}
\expandafter\def\csname Tie:Q1_2026:D30_AVAILABILITY_B10:EPSS_OPERATIONAL:r\endcsname{7}
\expandafter\def\csname Tie:Q1_2026:D30_AVAILABILITY_B10:EPSS_OPERATIONAL:a\endcsname{3}
\expandafter\def\csname Tie:Q1_2026:D30_AVAILABILITY_B10:EPSS_OPERATIONAL:q\endcsname{0}
\expandafter\def\csname Tie:Q1_2026:D30_AVAILABILITY_B10:EPSS_OPERATIONAL:k\endcsname{1,515}
\expandafter\def\csname Tie:Q1_2026:D30_AVAILABILITY_B10:EPSS_OPERATIONAL:minimum\endcsname{3}
\expandafter\def\csname Tie:Q1_2026:D30_AVAILABILITY_B10:EPSS_OPERATIONAL:maximum\endcsname{3}
\expandafter\def\csname TieMechanism:Q1_2026:cutoff_score\endcsname{9.8}
\expandafter\def\csname TieMechanism:Q1_2026:A\endcsname{169}
\expandafter\def\csname TieMechanism:Q1_2026:T\endcsname{1,434}
\expandafter\def\csname TieMechanism:Q1_2026:r\endcsname{589}
\expandafter\def\csname TieMechanism:Q1_2026:a\endcsname{1}
\expandafter\def\csname TieMechanism:Q1_2026:q\endcsname{4}
\expandafter\def\csname TieMechanism:Q1_2026:tied_positive_selected\endcsname{3}
\expandafter\def\csname TieMechanism:Q1_2026:tied_positive_not_selected\endcsname{1}
\expandafter\def\csname TieMechanism:Q1_2026:last_selected_date\endcsname{11 February 2026}
\expandafter\def\csname TieMechanism:Q1_2026:last_selected_timestamp\endcsname{2026-02-11T21:16:10.617000Z}
\expandafter\def\csname TieMechanism:Q1_2026:last_selected_cve\endcsname{CVE-2020-37176}
\expandafter\def\csname TieMechanism:Q1_2026:first_not_selected_date\endcsname{11 February 2026}
\expandafter\def\csname TieMechanism:Q1_2026:first_not_selected_timestamp\endcsname{2026-02-11T21:16:11.503000Z}
\expandafter\def\csname TieMechanism:Q1_2026:first_not_selected_cve\endcsname{CVE-2020-37181}
\expandafter\def\csname TieMechanism:Q1_2026:block1_above\endcsname{77}
\expandafter\def\csname TieMechanism:Q1_2026:block1_tied_selected\endcsname{589}
\expandafter\def\csname TieMechanism:Q1_2026:block2_above\endcsname{92}
\expandafter\def\csname TieMechanism:Q1_2026:block2_tied_selected\endcsname{0}
\expandafter\def\csname TieMechanism:Q1_2026:positive1_date\endcsname{23 January 2026}
\expandafter\def\csname TieMechanism:Q1_2026:positive1_cve\endcsname{CVE-2026-0770}
\expandafter\def\csname TieMechanism:Q1_2026:positive2_date\endcsname{29 January 2026}
\expandafter\def\csname TieMechanism:Q1_2026:positive2_cve\endcsname{CVE-2026-1340}
\expandafter\def\csname TieMechanism:Q1_2026:positive3_date\endcsname{6 February 2026}
\expandafter\def\csname TieMechanism:Q1_2026:positive3_cve\endcsname{CVE-2026-21643}
\expandafter\def\csname TieMechanism:Q1_2026:positive4_date\endcsname{11 March 2026}
\expandafter\def\csname TieMechanism:Q1_2026:positive4_cve\endcsname{CVE-2025-67038}
\expandafter\def\csname Case:68613:cve_id\endcsname{CVE-2025-68613}
\expandafter\def\csname Case:68613:publication_date\endcsname{19 December 2025}
\expandafter\def\csname Case:68613:kev_date\endcsname{11 March 2026}
\expandafter\def\csname Case:68613:d7_decision_date\endcsname{26 December 2025}
\expandafter\def\csname Case:68613:d7_cvss\endcsname{9.9}
\expandafter\def\csname Case:68613:d7_epss_display\endcsname{0.035}
\expandafter\def\csname Case:68613:d30_decision_date\endcsname{18 January 2026}
\expandafter\def\csname Case:68613:d30_cvss\endcsname{8.8}
\expandafter\def\csname Case:68613:d30_epss_display\endcsname{0.638}
\expandafter\def\csname TieBlock:2025:selected_tied_positives\endcsname{2}
\expandafter\def\csname TieBlock:2025:block_end_rank\endcsname{3,984}
\expandafter\def\csname TieBlock:Q1_2026:selected_tied_positives\endcsname{3}
\expandafter\def\csname TieBlock:Q1_2026:block_end_rank\endcsname{1,603}

\newsavebox{\articletablebox}
\newcommand{\wideinput}[1]{%
  \begingroup\small\setlength{\tabcolsep}{4pt}%
  \sbox{\articletablebox}{#1}%
  \ifdim\wd\articletablebox>\linewidth
    \resizebox{\linewidth}{!}{\usebox{\articletablebox}}%
  \else
    \usebox{\articletablebox}%
  \fi
  \endgroup}

\newcommand{\fullwidthinput}[1]{%
  \begingroup\small\setlength{\tabcolsep}{4pt}%
  \renewenvironment{tabular}[1]{%
    \begin{tabular*}{\linewidth}{@{\extracolsep{\fill}}##1@{}}%
  }{\end{tabular*}}%
  #1%
  \endgroup}

\makeatletter
\newenvironment{inlinetable}{%
  \let\@vspace\@vspace@orig
  \let\@vspacer\@vspacer@orig
  \noindent\begin{minipage}{\linewidth}\captionsetup{type=table}%
}{%
  \end{minipage}\par\medskip
}
\makeatother

\AtBeginEnvironment{thebibliography}{\interlinepenalty=10000\relax}

\input{sections/author_typography}
\title{Ties at the Top: A Point-in-Time Comparison of CVSS and EPSS at a Fixed Budget}
\author{Gianluca Martinuz}
\orcid{0009-0005-5696-2182}
\author{Paolo Ceravolo}
\orcid{0000-0002-4519-0173}
\affiliation{\institution{Università degli Studi di Milano}\city{Milan}\country{Italy}}
\authorsaddresses{}
\renewcommand{\shortauthors}{Martinuz and Ceravolo}
\hypersetup{pdfauthor={Gianluca Martinuz, Paolo Ceravolo},pdftitle={Ties at the Top: A Point-in-Time Comparison of CVSS and EPSS at a Fixed Budget}}
\begin{document}
\begin{abstract}
Organizations must decide which vulnerabilities warrant further analysis and protective action when the volume of published vulnerabilities exceeds their analytical capacity. We present a methodology for comparing vulnerability-prioritization strategies under an explicit resource constraint, represented by a fixed number of CVEs selected from a common eligible population. We apply this methodology to the Common Vulnerability Scoring System (CVSS) and the Exploit Prediction Scoring System (EPSS). For CVEs published in the National Vulnerability Database (NVD) from 17 March to 31 December 2025 and from 1 January to 31 March 2026 (Q1), we reconstruct each CVE's scores and status \param{primary_horizon} days after NVD publication, using archived source data. Each cohort has one ranking per strategy at a \param{primary_budget_pct}\% budget. Positives are CVEs added to the Known Exploited Vulnerabilities (KEV) catalog within \param{outcome_days} days after their decision day. The catalog is maintained by the Cybersecurity and Infrastructure Security Agency (CISA). EPSS captures \val{2025}{D30_AVAILABILITY_B05}{EPSS_OPERATIONAL}{captured} of \val{2025}{D30_AVAILABILITY_B05}{COMMON}{P} positives in 2025, versus \val{2025}{D30_AVAILABILITY_B05}{CVSS_OPERATIONAL}{captured} for CVSS; alternative selections among tied CVEs cannot raise CVSS capture above \val{2025}{D30_AVAILABILITY_B05}{CVSS_OPERATIONAL}{maximum}. In Q1, CVSS captures \val{Q1_2026}{D30_AVAILABILITY_B05}{CVSS_OPERATIONAL}{captured} of \val{Q1_2026}{D30_AVAILABILITY_B05}{COMMON}{P} and EPSS \val{Q1_2026}{D30_AVAILABILITY_B05}{EPSS_OPERATIONAL}{captured}, but tie-breaking can reverse this ordering. The score alone leaves most of each CVSS list undecided: the tie-breaking rule allocates \tie{2025}{D30_AVAILABILITY_B05}{CVSS_OPERATIONAL}{r} of \val{2025}{D30_AVAILABILITY_B05}{COMMON}{k} CVSS places in 2025 and \tie{Q1_2026}{D30_AVAILABILITY_B05}{CVSS_OPERATIONAL}{r} of \val{Q1_2026}{D30_AVAILABILITY_B05}{COMMON}{k} in Q1 among CVEs tied at score \tie{2025}{D30_AVAILABILITY_B05}{CVSS_OPERATIONAL}{cutoff_score}. Most chosen CVSS scores come from non-NVD providers; disabling those scores reduces CVSS capture from \val{2025}{D30_AVAILABILITY_B05}{CVSS_OPERATIONAL}{captured} to \val{2025}{D30_NVD_SOURCE_B05}{CVSS_OPERATIONAL}{captured} in 2025 and from \val{Q1_2026}{D30_AVAILABILITY_B05}{CVSS_OPERATIONAL}{captured} to \val{Q1_2026}{D30_NVD_SOURCE_B05}{CVSS_OPERATIONAL}{captured} in Q1. A fixed-budget comparison should therefore report its decision cutoff, eligible population, score provenance, and tie-breaking rule.
\end{abstract}
\maketitle
\raggedbottom
\hypersetup{pdfauthor={Gianluca Martinuz, Paolo Ceravolo}}
\section{Introduction}
Organizations face a large and continually changing set of vulnerabilities that may expose their systems to attack. Deciding where to intervene requires analytical work to assess relevance, investigate exploitation evidence, and identify appropriate protective measures. Limited analyst time and supporting resources constrain the number of vulnerabilities that can receive this attention, making prioritization a resource-allocation problem.

We present a methodology for evaluating alternative prioritization strategies within an explicit analytical budget. In this study, we represent that budget as a fixed number of CVEs selected for further assessment, providing a common basis for comparing strategies using information available at a defined decision time. This approach makes it possible to examine how score availability, provenance, and tie-breaking affect which CVEs are selected.

The Common Vulnerability Scoring System (CVSS) describes vulnerability characteristics and severity~\cite{firstcvss31}. The Exploit Prediction Scoring System (EPSS) estimates near-term exploitation probability~\cite{firstepss}. We compare their rankings against subsequent listing in the Known Exploited Vulnerabilities (KEV) catalog of the Cybersecurity and Infrastructure Security Agency (CISA). Our question is: \emph{At a fixed decision horizon and selection budget, how do CVSS and EPSS compare in capturing subsequent KEV listings, and how do score provenance and cutoff ties affect that comparison?}

CVSS vectors, EPSS scores, and National Vulnerability Database (NVD) status can change after publication. We reconstruct them at fixed horizons after NVD publication, using archived source data and keeping later outcomes separate from decision inputs.

A resource constraint makes the selection boundary consequential. When more CVEs share the cutoff score than the remaining places can accommodate, the score alone cannot determine which of them are selected. We call this group the \emph{cutoff tie block}. The tie-breaking rule then becomes part of the resource-allocation policy, and its effect must be included in the comparison. In our primary comparison, \param{primary_horizon} days after NVD publication at a \param{primary_budget_pct}\% budget, both CVSS cutoffs fall at score \tie{2025}{D30_AVAILABILITY_B05}{CVSS_OPERATIONAL}{cutoff_score}. The tie-breaking rule assigns \tie{2025}{D30_AVAILABILITY_B05}{CVSS_OPERATIONAL}{r} of \val{2025}{D30_AVAILABILITY_B05}{COMMON}{k} CVSS places in 2025 and \tie{Q1_2026}{D30_AVAILABILITY_B05}{CVSS_OPERATIONAL}{r} of \val{Q1_2026}{D30_AVAILABILITY_B05}{COMMON}{k} in Q1 2026. The score alone therefore leaves most of each CVSS list undecided. We report capture under the fixed rule and its minimum and maximum across alternative selections within this block. The 2025 EPSS advantage survives all these CVSS alternatives, whereas the Q1 ordering can reverse.

CVE-2025-68613 illustrates why timing and provenance also matter. Published on \gval{Case:68613:publication_date}, it had a reconstructed CVSS score of \gval{Case:68613:d7_cvss} from a non-NVD provider and an EPSS score of \gval{Case:68613:d7_epss_display} at D+\param{sensitivity_horizon} (\param{sensitivity_horizon} days after publication), when both strategies selected it at the \param{primary_budget_pct}\% budget. At D+\param{primary_horizon}, our fixed provider precedence selected an NVD vector with a Base score of \gval{Case:68613:d30_cvss}, placing the CVE below the CVSS cutoff. EPSS had risen to \gval{Case:68613:d30_epss_display}, keeping it in the EPSS list. The CVE was listed in KEV on \gval{Case:68613:kev_date}.

The allocation problem persists in automated vulnerability analysis and patch generation, as illustrated by DARPA's AI Cyber Challenge~\cite{darpa2025}. Analysis, testing, and review still consume resources. Our experiment evaluates CVE selection and subsequent KEV listing; it does not measure token consumption, remediation savings, or patching performance.

We make three contributions:\nopagebreak[4]
\begin{enumerate}
\item We provide a methodology for comparing vulnerability-prioritization strategies under a fixed selection budget, combining point-in-time reconstruction, common eligible populations, documented score provenance, and explicit analysis of cutoff ties, and apply it to two publication cohorts.
\item We show which capture differences survive alternative selections within cutoff tie blocks, alongside a separately identified post hoc uniform-selection reference.
\item We document score provenance and compare NVD-vector availability by subsequent KEV outcome, with supporting records for individual positive CVEs.
\end{enumerate}

\section{Background and related work}
\subsection{Severity, exploitation likelihood, and catalog evidence}
CVSS Base metrics describe intrinsic vulnerability properties rather than the probability of exploitation in a specified future period~\cite{firstcvss31}. NVD distributes its own vectors and those supplied by other providers: the repository and score provider are distinct roles. Wunder et al. documented inconsistencies in CVSS v3.1 assessments across evaluators~\cite{wunder2024}, reinforcing the need to record score provenance.

EPSS estimates exploitation likelihood~\cite{jacobs2021}, publishing daily probabilities of exploitation in the wild over the next 30 days~\cite{firstepss}. Jacobs et al. assess predictions through the share of relevant vulnerabilities captured and the effort spent on selections~\cite{jacobs2023}. We use related allocation measures, but define positives as subsequent KEV listings.

KEV records known exploited vulnerabilities and their listing dates~\cite{cisakev,cisakevschema}. The listing date is reproducible but does not identify exploitation onset. Entries already listed at decision time are excluded from our future-listing comparison.

\subsection{Temporal and allocation choices in related studies}
Allodi and Massacci assess CVSS-based remediation policies with a case-control design~\cite{allodi2014}. Jacobs et al. compare remediation strategies through coverage and efficiency against observed exploitation~\cite{jacobs2020}. Our outcome is subsequent KEV listing after an individual decision cutoff.

Koscinski et al. compare scoring systems under several evaluation designs~\cite{koscinski2025}. Their Microsoft corpus uses EPSS at a fixed date. A separate KEV analysis uses the maximum EPSS among monthly snapshots before catalog entry. Neither observes scores at the same fixed age for every CVE. Their top-$N$ selection explicitly expands to include ties. We instead retain exactly $k$ CVEs and quantify the capture range within the cutoff tie block. The two rules therefore answer different allocation questions.

Shimizu and Hashimoto evaluate prioritization against vendor reports of exploitation, using EPSS and KEV snapshots at a common date~\cite{shimizu2025}. KEV serves as an input and baseline, whereas our outcome is a later entry after each CVE's decision cutoff. Their efficiency and coverage measures relate to selected-list precision and recall, but timing and labels prevent direct comparison of capture rates.

Our comparison combines individual cutoffs, exclusion of already-known KEV entries, an exact common budget, and explicit score provenance. Applied to the same reconstructed data, these choices show which capture differences survive alternative selections among tied CVEs and how score coverage affects the evaluated lists.

\section{Data and temporal reconstruction}
\subsection{Cohorts, sources, and acquisition chronology}
The 2025 cohort covers CVEs published in NVD from 17 March through 31 December 2025; its start coincides with the first published EPSS v4 scores (model identifier \texttt{\source{epss_model}})~\cite{epssarchive}. The first quarter of 2026 (Q1) covers 1 January through 31 March. Membership follows NVD publication, not the year in the CVE identifier. We evaluate the cohorts separately.

For 2025, we recomputed the earlier results~\cite{thesis2026} with the corrected reconstruction in Section~\ref{sec:corrections}. Population counts and some selections changed; primary captured-positive counts did not. For Q1, we use the same primary comparison for a later period.

NVD records and change-history events supply publication timestamps, status, and CVSS vectors~\cite{nvdcveapi,nvdhistoryapi}. Archived daily FIRST files supply EPSS scores~\cite{epssarchive}; KEV supplies listing dates~\cite{cisakev}.

We retrieved NVD data and KEV release \texttt{\source{kev_release_2025}} for the 2025 cohort on \source{nvd_retrieved_2025}, and new NVD data and KEV release \texttt{\source{kev_release_Q1}} for Q1 on \source{kev_retrieved_Q1}. The final D+\param{primary_horizon} outcome windows end on 29 July 2026 for the 2025 cohort and 27 October 2026 for Q1, before their respective acquisitions. Comparing the catalog retrieved on 2 August, release \texttt{\source{kev_release_2025}}, with the November release found no additional entries dated on or before 29 July 2026 and no changes to existing listing dates.

We retained \source{epss_snapshots_2025} daily EPSS snapshots for the 2025 cohort and \source{epss_snapshots_Q1_retained} for Q1 2026. All \source{epss_snapshots_2025} snapshots for 2025 and \source{epss_snapshots_Q1_used} of the \source{epss_snapshots_Q1_retained} for Q1 supplied scores at the observed decision dates; no CVEs in the Q1 cohort had decision dates on the remaining two days. All archived EPSS files record model identifier \texttt{\source{epss_model}}. FIRST began publishing EPSS v5 on 15 June 2026, after the last decision day of both cohorts~\cite{epssarchive}.

\subsection{Decision time and reconstructed information}
\begin{figure}[!htbp]
\centering\includegraphics[width=\linewidth]{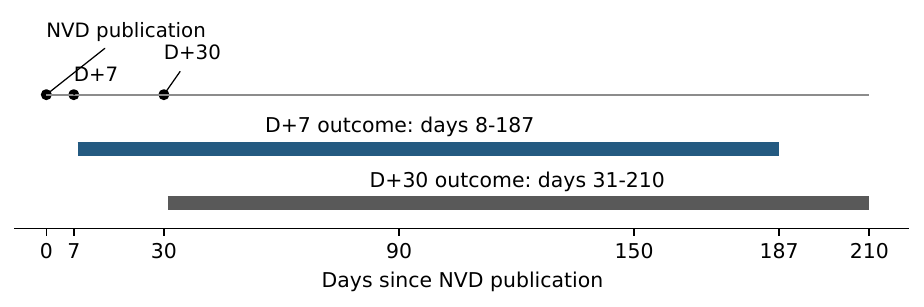}
\caption{Decision horizons and subsequent KEV windows relative to NVD publication. Each cutoff has its own reconstructed scores and eligibility assessment. Later KEV listings are outcomes, not ranking inputs.}
\Description{Timeline separating NVD publication, the seven-day and thirty-day decision cutoffs, and their subsequent KEV windows.}
\label{fig:timeline}
\end{figure}

For publication on UTC calendar day $p$ and horizon $h\in\{\param{sensitivity_horizon},\param{primary_horizon}\}$, the decision day is $d_h=p+h$, with its cutoff at the end of that UTC day. Ranking ties use the full NVD publication timestamp; decision and outcome calculations use its calendar date (Figure~\ref{fig:timeline}).

Using historical data acquired later, we reconstruct NVD status and each CVSS source/version pair from events no later than the cutoff. Additions, changes, and removals apply to the corresponding pair. We do not carry an earlier value forward through an uninterpretable event. Simultaneous events are accepted only if all compatible event orderings yield the same state. A later rejection or reinstatement does not alter the state at an earlier cutoff.

Usable vectors follow fixed precedence: NVD v3.1, non-NVD v3.1, NVD v3.0, then non-NVD v3.0. Within each CVSS version, non-NVD providers are ordered by source identifier. We recalculate the Base score from the chosen vector and retain its provider and version. CVSS v2.0 and v4.0 are neither used nor converted. Non-NVD providers include CVE Numbering Authorities (CNAs) and Authorized Data Publishers (ADPs).

For EPSS, we require a daily archive covering the full decision-date interval; a gap stops processing. We select the latest valid snapshot dated no later than each CVE's decision day, which is the snapshot of that day in both cohorts. An absent or unresolved CVE record leaves EPSS unusable; we do not search older snapshots for a replacement. Ranking uses probability, not percentile.

\subsection{Eligibility, coverage, and the Q1 extension}
A CVE is eligible if it has determinable dates, a complete outcome window, a resolvable non-rejected NVD state, no KEV listing known at the cutoff, and at least one usable score. An unresolved state excludes the CVE; an unresolved score disables only that score if the other is usable. Exclusion reasons follow fixed precedence, so the exclusion counts in Table~\ref{tab:dataset} do not overlap.

Table~\ref{tab:dataset} shows cohort flow and coverage. No further exclusions follow the status and known-KEV checks. All eligible CVEs have EPSS; complete cases also have usable CVSS.

\begin{inlinetable}
\caption{Cohort flow and score coverage. Scope: CVEs before exclusions; $N$: eligible CVEs; $P$: eligible positives in the corresponding outcome window. Exclusion reasons follow precedence order. Residual exclusions and eligible CVSS-only counts are zero throughout.}
\label{tab:dataset}
\centering\wideinput{
\begin{tabular}{llrrrrrrr}
\toprule
Cohort & Horizon & Scope & Rejected & Known KEV & $N$ & Complete cases & EPSS only & $P$ \\
\midrule
2025 & D+7 & 39,917 & 1,371 & 70 & 38,476 & 30,741 & 7,735 & 60 \\
2025 & D+30 & 39,917 & 1,457 & 92 & 38,368 & 31,618 & 6,750 & 40 \\
Q1 2026 & D+7 & 16,255 & 1,068 & 27 & 15,160 & 13,258 & 1,902 & 16 \\
Q1 2026 & D+30 & 16,255 & 1,081 & 33 & 15,141 & 13,934 & 1,207 & 11 \\
\bottomrule
\end{tabular}
}
\end{inlinetable}

A preliminary exploratory analysis covered CVEs published from 1 January through 17 February 2026. It was not validated; its figures are not reported here. The period was extended to 31 March with the same primary comparison criteria. The specification dated 1 November 2026 predates acquisition of the extension data. Q1 results were derived anew from retained sources, without importing exploratory result tables.

Each Q1 comparison uses one ranking and budget per strategy for the full quarter. The publication periods 1 January--17 February and 18 February--31 March provide descriptive detail within those lists.

\section{Evaluation design}\label{sec:design}
\subsection{A common population and an exact selection budget}
An evaluation cell fixes cohort, horizon, budget, and population and score filters. Both strategies use the same eligible population of size $N$ and select exactly $k=\lceil bN\rceil$ CVEs, where $b$ is the budget fraction. The primary cell uses D+\param{primary_horizon} and \param{primary_budget_pct}\%. The budget is recalculated from the population remaining after each sensitivity filter. Equal budgets mean equal CVE counts, not equal analysis or remediation costs.

CVSS ranks usable Base scores in descending order; EPSS ranks probabilities the same way. Unusable scores are ranked after all usable ones; no zero is imputed. Ties use ascending NVD publication timestamp, then ascending CVE identifier. This deterministic rule was specified before the primary analysis and uses fields available at the cutoff. No additional rounding is applied: CVSS Base scores keep the rounding required by the specification, and EPSS probabilities are used as published. If selection reaches unusable scores, we apply the same secondary keys and record insufficient coverage. This does not occur here.

Rankings aggregate scores at the same age after NVD publication. They are not a queue available on one historical date and do not simulate arrivals, patch completion, or continuous reprioritization. Table~\ref{tab:method} summarizes the key choices in our methodology and provides a reporting checklist: its decision categories are reusable, but settings remain study-specific.
\begin{table}[!htbp]
\caption{Essential rules of the point-in-time, fixed-budget comparison.}
\label{tab:method}
\centering\small
\begin{tabular}{p{0.15\linewidth}p{0.77\linewidth}}
\toprule
Decision & Rule \\
\midrule
Cutoff & End of day D+30 UTC; D+7 as sensitivity. \\
Information & Reconstruct scores and NVD status from retained source records. \\
Eligibility & Resolve status, exclude known KEV, require at least one usable score. \\
Population & Use the same eligible CVEs for both strategies within each cell. \\
Budget & $k=\lceil bN\rceil$; primary $b=5\%$, sensitivities $1\%$ and $10\%$. \\
Ordering & Descending score; ascending NVD publication time, then CVE identifier. \\
Outcome & KEV entry after the decision day and within the next 180 days. \\
Ties & Keep exactly $k$; report observed capture and combinatorial bounds. \\
Checks & Verify inputs, selections, counts, and provenance; retain corrections. \\
\bottomrule
\end{tabular}

\end{table}

\subsection{Outcome and comparison measures}
A CVE is a positive if its KEV listing date falls after the decision day and no later than \param{outcome_days} days after it. Same-day listings are already known and excluded before ranking. Missing, invalid, or conflicting dates in KEV entries halt processing. An outcome of zero denotes no documented listing within the outcome window, not absence of exploitation.

Recall is the selected share of the $P$ eligible positives; yield is the positive share of the $k$ selected CVEs, or selected-list precision. These allocation measures follow the efficiency/coverage perspective~\cite{jacobs2020,jacobs2021,jacobs2023}, with KEV listing as the outcome here. Counts and denominators accompany the rates; a zero denominator yields an undefined rate.

If the lists share $I$ CVEs, each contains $k-I$ CVEs absent from the other; their Jaccard index is $I/(2k-I)$. We also count positives selected by both strategies, CVSS only, EPSS only, or neither.

\subsection{What a cutoff tie permits}\label{sec:ties}
The fixed tie-breaking result remains primary. Alternatives at the same budget vary selection only within the cutoff block, keeping CVEs above it selected and those below it unselected. Let $A$ be the number of CVEs above the block and $a$ their positive count, $T$ the number of tied CVEs and $q$ their positive count, and $r=k-A$ the number of places left to fill. Attainable capture is bounded by
\[
C_{\min}=a+\max\{0,r-(T-q)\},\qquad
C_{\max}=a+\min\{r,q\}.
\]
Outcomes enter this diagnostic after selection. These are combinatorial bounds, not confidence intervals or estimates of variation across populations.

After observing the primary results, we added the expected capture under uniform sampling of $r$ CVEs without replacement from the $T$ tied CVEs: $a+rq/T$. This expectation weights tied CVEs equally and need not equal the midpoint of the bounds. We report it for both strategies in both primary cells. Exact fractions are included in the ancillary files; displayed values use decimal half-up rounding to one decimal. The diagnostic does not replace the selection rule or test its bias.

\subsection{Sensitivity, stability, and descriptive diagnostics}\label{sec:sensitivity}
Planned sensitivities vary one element at a time. D+\param{primary_horizon} budgets of \param{lower_budget_pct}\% and \param{upper_budget_pct}\% retain the primary population and ordering. Complete cases require both scores. The NVD-source variant disables the previously chosen CVSS score if its provider is non-NVD, without retrieving an alternative vector. The v3.1-only variant disables a chosen v3.0 score. For these two variants, we retain CVEs with at least one usable score and recalculate population and budget.

D+\param{sensitivity_horizon} repeats reconstruction and evaluation at \param{primary_budget_pct}\%. Its outcome covers days 8--187 after NVD publication, versus 31--210 at D+\param{primary_horizon}; eligibility and labels are recalculated. For stability, we restrict both rankings to the $N_{\cap}$ CVEs eligible at both horizons and apply the same budget rule at \param{primary_budget_pct}\%, obtaining $k_{\cap}$. Turnover is the number of CVEs entering the D+\param{primary_horizon} list divided by $k_{\cap}$; with equal list sizes it also equals the fraction leaving. Absolute rank movement is the absolute difference between the two ranks divided by $N_{\cap}-1$.

A second post hoc diagnostic identifies positives shared by both horizons or found at only one, and documents the reasons for differences in positive-set membership. It also describes D+\param{sensitivity_horizon} selection among CVEs added to KEV 8--30 days after NVD publication.

Chosen-score provenance groups eligible CVEs by their chosen CVSS provider, separating those without usable CVSS. NVD availability instead records any usable NVD v3.0 or v3.1 vector at D+\param{primary_horizon}, regardless of the chosen score. Its share within each outcome group is available/(available+absent), with unresolved records separate.

\section{Results}
\subsection{Primary comparison and capture bounds}
Each strategy selects \val{2025}{D30_AVAILABILITY_B05}{COMMON}{k} of \val{2025}{D30_AVAILABILITY_B05}{COMMON}{N} eligible CVEs in 2025 and \val{Q1_2026}{D30_AVAILABILITY_B05}{COMMON}{k} of \val{Q1_2026}{D30_AVAILABILITY_B05}{COMMON}{N} in Q1 2026. Subsequent KEV positives represent \val{2025}{D30_AVAILABILITY_B05}{COMMON}{base_rate} and \val{Q1_2026}{D30_AVAILABILITY_B05}{COMMON}{base_rate} of these populations, respectively.

CVSS captures \val{2025}{D30_AVAILABILITY_B05}{CVSS_OPERATIONAL}{captured} of \val{2025}{D30_AVAILABILITY_B05}{COMMON}{P} positives in 2025 and EPSS captures \val{2025}{D30_AVAILABILITY_B05}{EPSS_OPERATIONAL}{captured}. In Q1, CVSS captures \val{Q1_2026}{D30_AVAILABILITY_B05}{CVSS_OPERATIONAL}{captured} of \val{Q1_2026}{D30_AVAILABILITY_B05}{COMMON}{P}, while EPSS captures \val{Q1_2026}{D30_AVAILABILITY_B05}{EPSS_OPERATIONAL}{captured} (Table~\ref{tab:primary}). At this \param{primary_budget_pct}\% budget, \val{2025}{D30_AVAILABILITY_B05}{COMMON}{neither} of \val{2025}{D30_AVAILABILITY_B05}{COMMON}{P} positives in 2025 and \val{Q1_2026}{D30_AVAILABILITY_B05}{COMMON}{neither} of \val{Q1_2026}{D30_AVAILABILITY_B05}{COMMON}{P} in Q1 lie outside both selected lists.

\begin{table}[!htbp]
\caption{Primary D+\param{primary_horizon} comparison at a \param{primary_budget_pct}\% budget. The bounds reflect alternative selections within the cutoff tie block; they are not confidence intervals. Composition columns count positives, whereas list overlap and Jaccard compare all selected CVEs. The final panel reports the post hoc uniform-selection reference.}
\label{tab:primary}
\centering\begin{minipage}{0.90\linewidth}
\small
\textbf{(a) Primary results}\par\smallskip
\fullwidthinput{
\begin{tabular}{lrrrrrrr}
\toprule
Strategy & $N$ & $k$ & $P$ & Captured & Recall & Yield & \shortstack[r]{Capture\\bounds} \\
\midrule
\multicolumn{8}{l}{\textbf{2025 cohort}} \\
CVSS & 38,368 & 1,919 & 40 & 5 & 12.5\% & 0.26\% & [3, 9] \\
EPSS & 38,368 & 1,919 & 40 & 16 & 40.0\% & 0.83\% & [16, 16] \\
\multicolumn{8}{l}{\textbf{Q1 2026 cohort}} \\
CVSS & 15,141 & 758 & 11 & 4 & 36.4\% & 0.53\% & [1, 5] \\
EPSS & 15,141 & 758 & 11 & 3 & 27.3\% & 0.40\% & [3, 3] \\
\bottomrule
\end{tabular}
}
\medskip\par\textbf{(b) Positive composition and list overlap}\par\smallskip
\fullwidthinput{
\begin{tabular}{lrrrrrr}
\toprule
Cohort & Both & CVSS only & EPSS only & Neither & List overlap & Jaccard \\
\midrule
2025 & 2 & 3 & 14 & 21 & 268 & 0.075 \\
Q1 2026 & 2 & 2 & 1 & 6 & 135 & 0.098 \\
\bottomrule
\end{tabular}
}
\medskip\par\textbf{(c) Post hoc uniform tie-breaking reference}\par\smallskip
\fullwidthinput{
\begin{tabular}{llrrrrrrr}
\toprule
Cohort & Strategy & Observed & \shortstack[r]{Capture\\bounds} & \shortstack[r]{Uniform\\mean} & $T$ & $r$ & $a$ & $q$ \\
\midrule
2025 & CVSS & 5 & [3, 9] & 5.7 & 3,743 & 1,678 & 3 & 6 \\
2025 & EPSS & 16 & [16, 16] & 16.0 & 7 & 3 & 16 & 0 \\
Q1 2026 & CVSS & 4 & [1, 5] & 2.6 & 1,434 & 589 & 1 & 4 \\
Q1 2026 & EPSS & 3 & [3, 3] & 3.0 & 5 & 4 & 3 & 0 \\
\bottomrule
\end{tabular}
}
\end{minipage}
\end{table}

\begin{figure}[!htbp]
\centering\includegraphics[width=\linewidth]{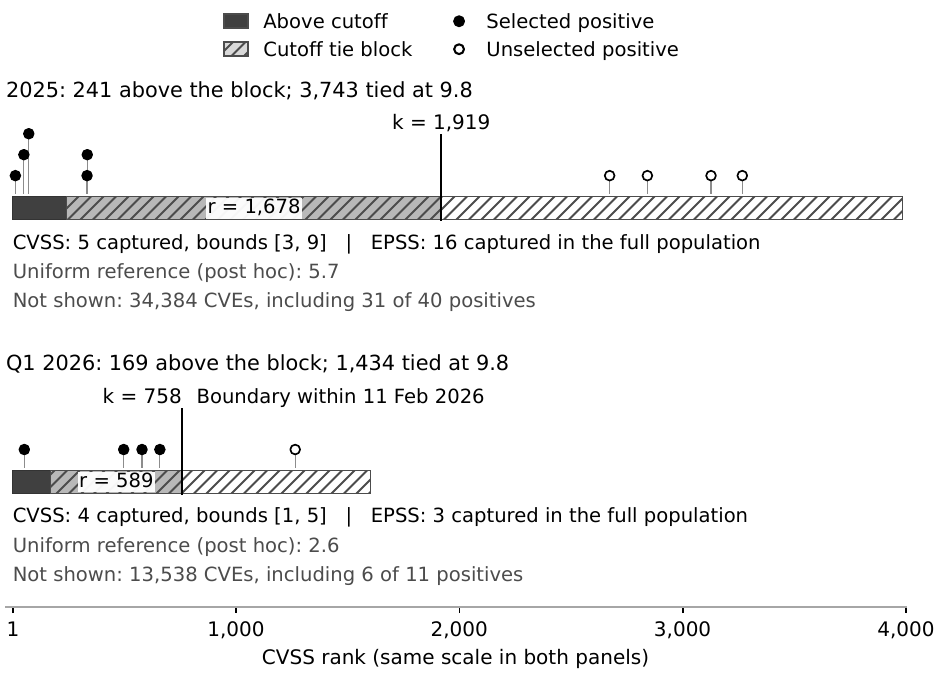}
\caption{The CVSS cutoff tie block in the primary comparison (D+\param{primary_horizon}, \param{primary_budget_pct}\%). Both panels use the same rank scale; each strip ends at the end of the cutoff block at score \tie{2025}{D30_AVAILABILITY_B05}{CVSS_OPERATIONAL}{cutoff_score}. Filled and open circles mark selected and unselected positives; vertical offsets separate nearby ranks. Within the hatched block, CVEs are ordered by NVD publication time and then CVE identifier; gray fill marks the $r$ selected CVEs. The vertical line marks the budget limit, falling among CVEs published on \mechanism{last_selected_date} in Q1. Positives below the block are counted but not drawn. EPSS captures refer to the whole eligible population at the same budget. Uniform references are post hoc expectations; bounds are combinatorial.}
\Description{Two CVSS ranking strips show the budget boundary within a large tie block. The 2025 EPSS capture exceeds the CVSS upper bound, whereas the Q1 EPSS capture lies within the CVSS bounds. Positive markers use their saved ranks, with vertical offsets for readability.}
\label{fig:cutoff-block}
\end{figure}

The capture bounds distinguish the two results (Figure~\ref{fig:cutoff-block}). In 2025, the maximum CVSS capture within its cutoff tie block is \val{2025}{D30_AVAILABILITY_B05}{CVSS_OPERATIONAL}{maximum}, below the \val{2025}{D30_AVAILABILITY_B05}{EPSS_OPERATIONAL}{captured} EPSS captures. That advantage survives all selections confined to the CVSS block. In Q1, the \val{Q1_2026}{D30_AVAILABILITY_B05}{EPSS_OPERATIONAL}{captured} EPSS captures lie within the CVSS~bounds~\mbox{[\val{Q1_2026}{D30_AVAILABILITY_B05}{CVSS_OPERATIONAL}{minimum}, \val{Q1_2026}{D30_AVAILABILITY_B05}{CVSS_OPERATIONAL}{maximum}]}, so alternative selections can reverse the ordering. Alternative selections within the EPSS cutoff blocks change list membership but not capture counts in either primary comparison. In 2025, publication-time ordering selects \gval{TieBlock:2025:selected_tied_positives} of the \tie{2025}{D30_AVAILABILITY_B05}{CVSS_OPERATIONAL}{q} positives in the CVSS cutoff tie block.

The retained Q1 ranking makes the mechanism explicit. Above the cutoff score \mechanism{cutoff_score} are \mechanism{A} CVEs, including \mechanism{a} positive. The tie block contains \mechanism{T} CVEs, including \mechanism{q} positives, for \mechanism{r} remaining places. Ascending publication-time ordering reaches \mechanism{last_selected_date}, with the boundary falling within that day. The \mechanism{tied_positive_selected} tied positives published before that date are selected; the remaining positive, published on \mechanism{positive4_date}, is not. CVSS thus captures \val{Q1_2026}{D30_AVAILABILITY_B05}{CVSS_OPERATIONAL}{captured} positives: \mechanism{a} above the block and \mechanism{tied_positive_selected} within it.

Uniform selection instead gives \val{Q1_2026}{D30_AVAILABILITY_B05}{CVSS_OPERATIONAL}{expected} expected captures. The ancillary files retain the cutoff boundary and positive-level evidence.

Across the two Q1 publication periods (1 January--17 February and 18 February--31 March), CVSS selects \block{Q1_2026}{1}{selected_cvss} and \block{Q1_2026}{2}{selected_cvss} CVEs, respectively, compared with \block{Q1_2026}{1}{selected_epss} and \block{Q1_2026}{2}{selected_epss} for EPSS. All \mechanism{block1_tied_selected} selected CVEs within the tie block belong to the first period.

\subsection{Budget sensitivity and the structure of the ties}
\begin{figure}[!htbp]
\centering\includegraphics[width=\linewidth]{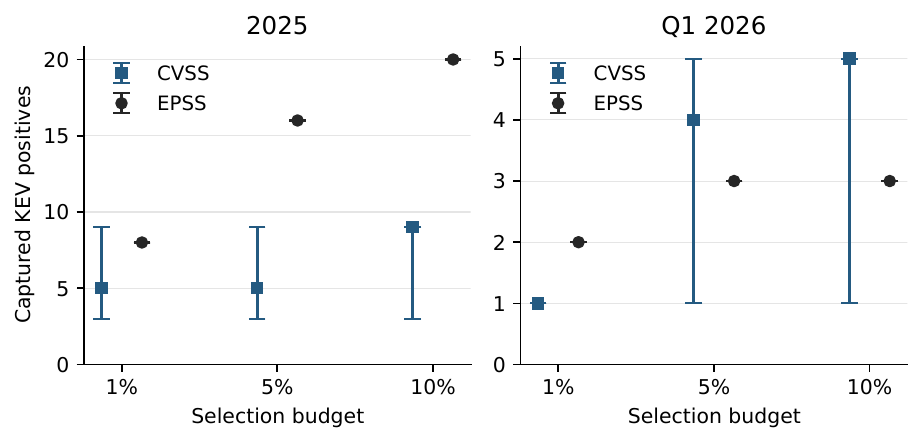}
\caption{Captured positives at the three evaluated D+\param{primary_horizon} budgets. Points denote the fixed tie-breaking result; segments show attainable minimum and maximum capture within the cutoff tie block. The two panels use different vertical scales. These segments are combinatorial bounds, and no interpolation between budgets is implied.}
\Description{Separate cohort panels comparing observed captures and combinatorial tie bounds at one, five, and ten percent budgets, with different vertical scales.}
\label{fig:budget}
\end{figure}

EPSS captures more positives at all three evaluated D+\param{primary_horizon} budgets in 2025 (Figure~\ref{fig:budget}). The lead survives cutoff-tie alternatives at \param{primary_budget_pct}\% and \param{upper_budget_pct}\%, but not at \param{lower_budget_pct}\%, where its \val{2025}{D30_AVAILABILITY_B01}{EPSS_OPERATIONAL}{captured} captures fall within the CVSS~bounds~\mbox{[\tie{2025}{D30_AVAILABILITY_B01}{CVSS_OPERATIONAL}{minimum}, \tie{2025}{D30_AVAILABILITY_B01}{CVSS_OPERATIONAL}{maximum}]}. All three CVSS budgets cut through the same block at score \tie{2025}{D30_AVAILABILITY_B05}{CVSS_OPERATIONAL}{cutoff_score}. All \tie{2025}{D30_AVAILABILITY_B05}{CVSS_OPERATIONAL}{q} positives in the block can fit either inside or outside each selection, leaving the bounds unchanged.

In Q1, EPSS leads at \param{lower_budget_pct}\% regardless of cutoff-tie alternatives; the ordering at \param{primary_budget_pct}\% and \param{upper_budget_pct}\% depends on tie-breaking. The two larger budgets share the block at \tie{Q1_2026}{D30_AVAILABILITY_B05}{CVSS_OPERATIONAL}{cutoff_score}. The \param{lower_budget_pct}\% budget instead cuts through a block at \tie{Q1_2026}{D30_AVAILABILITY_B01}{CVSS_OPERATIONAL}{cutoff_score}, containing \tie{Q1_2026}{D30_AVAILABILITY_B01}{CVSS_OPERATIONAL}{T} CVEs and no positives.

\subsection{Chosen-score provenance and NVD availability}
Non-NVD providers supply most chosen CVSS scores: \prov{2025}{CNA}{share} of eligible CVEs in 2025 and \prov{Q1_2026}{CNA}{share} in Q1 (Table~\ref{tab:provenance}). Q1 has a higher share of NVD-provided scores and a lower share without usable CVSS. Describing the comparison simply as using ``NVD data'' would conceal this mixture: NVD distributes and records the history of vectors assigned by multiple providers.

\begin{table}[!htbp]
\caption{Provenance of the CVSS score chosen for the primary ranking, with all eligible CVEs as denominator. Non-NVD providers include CNAs and ADPs. This classification differs from availability of any usable NVD vector.}
\label{tab:provenance}
\centering\wideinput{
\begin{tabular}{llrr}
\toprule
Cohort & Chosen-score provenance & CVEs & Share \\
\midrule
2025 & NVD & 7,507 & 19.6\% \\
2025 & Non-NVD providers & 24,111 & 62.8\% \\
2025 & No usable CVSS & 6,750 & 17.6\% \\
Q1 2026 & NVD & 4,897 & 32.3\% \\
Q1 2026 & Non-NVD providers & 9,037 & 59.7\% \\
Q1 2026 & No usable CVSS & 1,207 & 8.0\% \\
\bottomrule
\end{tabular}
}
\end{table}

\begin{figure}[!htbp]
\centering\includegraphics[width=\linewidth]{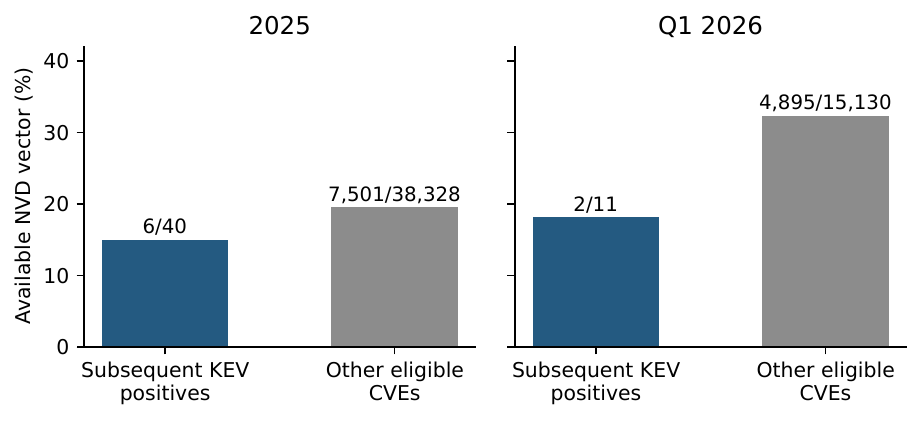}
\caption{NVD vector availability at D+\param{primary_horizon} by subsequent KEV outcome. Shares use available plus absent records as denominator. Resolved denominators are shown.}
\Description{NVD vector availability for subsequent KEV positives and other eligible CVEs in both cohorts, with counts and resolved denominators.}
\label{fig:nvd}
\end{figure}

NVD vector availability is lower among subsequent KEV positives than among other eligible CVEs in both cohorts (Figure~\ref{fig:nvd}): \nvd{2025}{positive}{available}/\nvd{2025}{positive}{denominator} (\nvd{2025}{positive}{share}) versus \nvd{2025}{negative}{available}/\nvd{2025}{negative}{denominator} (\nvd{2025}{negative}{share}) in 2025, and \nvd{Q1_2026}{positive}{available}/\nvd{Q1_2026}{positive}{denominator} (\nvd{Q1_2026}{positive}{share}) versus \nvd{Q1_2026}{negative}{available}/\nvd{Q1_2026}{negative}{denominator} (\nvd{Q1_2026}{negative}{share}) in Q1. Availability is higher in Q1 within each outcome group. These are descriptive associations; the Q1 comparison rests on only \val{Q1_2026}{D30_AVAILABILITY_B05}{COMMON}{P} positives. The ancillary files supply unresolved counts, coverage denominators, and monthly availability.

\subsection{Sensitivities, stability, and early KEV entries}
The observed EPSS lead in 2025 persists across all six planned sensitivities (Table~\ref{tab:sensitivity}), although it becomes tie-dependent at the \param{lower_budget_pct}\% budget. In Q1, EPSS exceeds the CVSS capture bounds at \param{lower_budget_pct}\% and under the NVD-source filter; the other variants have different observed orderings and overlapping capture bounds.

\begin{table}[!htbp]
\caption{Sensitivity comparisons. Each cell has its own population $N$, exact budget $k$, and positive count $P$. All rows use D+\param{primary_horizon} unless stated otherwise. Bounds reflect cutoff-tie alternatives, not confidence intervals. D+\param{sensitivity_horizon} also changes the outcome window. The primary comparison is in Table~\ref{tab:primary}; full metrics are included in the ancillary files.}
\label{tab:sensitivity}
\centering\wideinput{
\begin{tabular}{lrrrrrrr}
\toprule
Cell & $N$ & $k$ & $P$ & \shortstack[r]{CVSS\\captured} & \shortstack[r]{CVSS capture\\bounds} & \shortstack[r]{EPSS\\captured} & \shortstack[r]{EPSS capture\\bounds} \\
\midrule
\multicolumn{8}{l}{\textbf{2025 cohort}} \\
D+30, 1\% & 38,368 & 384 & 40 & 5 & [3, 9] & 8 & [8, 8] \\
D+30, 10\% & 38,368 & 3,837 & 40 & 9 & [3, 9] & 20 & [20, 20] \\
Complete cases, 5\% & 31,618 & 1,581 & 34 & 5 & [3, 9] & 12 & [12, 12] \\
NVD source filter, 5\% & 38,368 & 1,919 & 40 & 1 & [0, 1] & 16 & [16, 16] \\
CVSS v3.1, 5\% & 38,368 & 1,919 & 40 & 5 & [3, 9] & 16 & [16, 16] \\
D+7, 5\% & 38,476 & 1,924 & 60 & 14 & [8, 21] & 23 & [23, 24] \\
\multicolumn{8}{l}{\textbf{Q1 2026 cohort}} \\
D+30, 1\% & 15,141 & 152 & 11 & 1 & [1, 1] & 2 & [2, 2] \\
D+30, 10\% & 15,141 & 1,515 & 11 & 5 & [1, 5] & 3 & [3, 3] \\
Complete cases, 5\% & 13,934 & 697 & 11 & 4 & [1, 5] & 3 & [3, 3] \\
NVD source filter, 5\% & 15,141 & 758 & 11 & 0 & [0, 0] & 3 & [3, 3] \\
CVSS v3.1, 5\% & 15,141 & 758 & 11 & 3 & [1, 4] & 3 & [3, 3] \\
D+7, 5\% & 15,160 & 758 & 16 & 5 & [1, 6] & 6 & [6, 6] \\
\bottomrule
\end{tabular}
}
\end{table}

Complete cases form a smaller population, with fewer positives in 2025, so their capture counts are not directly comparable with the primary counts. Disabling the chosen non-NVD scores reduces CVSS capture to \val{2025}{D30_NVD_SOURCE_B05}{CVSS_OPERATIONAL}{captured} in 2025 and \val{Q1_2026}{D30_NVD_SOURCE_B05}{CVSS_OPERATIONAL}{captured} in Q1; this filter does not represent every NVD-only reconstruction. The v3.1 filter preserves the primary 2025 captures but changes Q1 CVSS capture to \val{Q1_2026}{D30_V31_ONLY_B05}{CVSS_OPERATIONAL}{captured}.

At D+\param{sensitivity_horizon}, the 2025 EPSS bounds \mbox{[\val{2025}{D7_AVAILABILITY_B05}{EPSS_OPERATIONAL}{minimum}, \val{2025}{D7_AVAILABILITY_B05}{EPSS_OPERATIONAL}{maximum}]} exceed the CVSS~bounds~\mbox{[\val{2025}{D7_AVAILABILITY_B05}{CVSS_OPERATIONAL}{minimum}, \val{2025}{D7_AVAILABILITY_B05}{CVSS_OPERATIONAL}{maximum}]}. In Q1, EPSS captures \val{Q1_2026}{D7_AVAILABILITY_B05}{EPSS_OPERATIONAL}{captured} positives against \val{Q1_2026}{D7_AVAILABILITY_B05}{CVSS_OPERATIONAL}{captured} for CVSS, whose upper bound is \val{Q1_2026}{D7_AVAILABILITY_B05}{CVSS_OPERATIONAL}{maximum}. Tie-breaking can erase, but not reverse, the Q1 EPSS lead at D+\param{sensitivity_horizon}. At D+\param{primary_horizon}, a different selection within the CVSS cutoff tie block can reverse the observed CVSS lead in Q1.

Among CVEs eligible at both D+\param{sensitivity_horizon} and D+\param{primary_horizon}, list turnover is lower for EPSS: \stability{2025}{EPSS_OPERATIONAL}{turnover} versus \stability{2025}{CVSS_OPERATIONAL}{turnover} for CVSS in 2025, and \stability{Q1_2026}{EPSS_OPERATIONAL}{turnover} versus \stability{Q1_2026}{CVSS_OPERATIONAL}{turnover} in Q1. The 90th percentile of absolute normalized rank movement is nevertheless higher for EPSS: \stability{2025}{EPSS_OPERATIONAL}{p90} versus \stability{2025}{CVSS_OPERATIONAL}{p90} for CVSS in 2025, and \stability{Q1_2026}{EPSS_OPERATIONAL}{p90} versus \stability{Q1_2026}{CVSS_OPERATIONAL}{p90} in Q1.

The post hoc reconciliation identifies \diag{2025}{common} positives shared across horizons in 2025 and \diag{Q1_2026}{common} in Q1. CVEs positive only at D+\param{sensitivity_horizon} number \diag{2025}{D7_only} and \diag{Q1_2026}{D7_only}, respectively, and entered KEV 8--30 days after NVD publication. The \diag{2025}{D30_only} and \diag{Q1_2026}{D30_only} positives found only at D+\param{primary_horizon} entered on days 188--210. Among the early entries, CVSS and EPSS each select \early{2025}{captured_cvss} of \early{2025}{subgroup_n} at D+\param{sensitivity_horizon} in 2025, including \early{2025}{both} shared selections. In Q1 they select \early{Q1_2026}{captured_cvss} and \early{Q1_2026}{captured_epss} of \early{Q1_2026}{subgroup_n}.
\FloatBarrier

\section{Discussion}
\subsection{The observed ordering and the role of ties}
The primary EPSS advantage in 2025 survives every selection within the CVSS cutoff tie block. In Q1, the choice among tied CVEs can reverse the ordering. Reporting observed captures alone would miss this distinction. At D+\param{primary_horizon}, this contrast holds at \param{primary_budget_pct}\% and \param{upper_budget_pct}\%. At \param{lower_budget_pct}\%, the EPSS lead in 2025 depends on tie-breaking, whereas the Q1 EPSS capture exceeds the CVSS upper bound (Figure~\ref{fig:budget}).

CVSS Base scores use one decimal place on a 0--10 scale, allowing at most 101 values~\cite{firstcvss31}. This limited resolution helps explain the large ties, whose size and composition also depend on cohort and budget. Both cohorts' CVSS lists end within a tie block at every evaluated D+\param{primary_horizon} budget. EPSS has finer resolution: its cutoff blocks contain between \tie{2025}{D30_AVAILABILITY_B01}{EPSS_OPERATIONAL}{T} and \tie{2025}{D30_AVAILABILITY_B10}{EPSS_OPERATIONAL}{T} CVEs.

When the budget cuts through a tie block, keeping an exact list size requires a choice among tied CVEs. At D+\param{primary_horizon} and the primary \param{primary_budget_pct}\% budget, including all CVEs tied at the cutoff would expand the CVSS lists from \val{2025}{D30_AVAILABILITY_B05}{COMMON}{k} to \gval{TieBlock:2025:block_end_rank} CVEs in 2025 and from \val{Q1_2026}{D30_AVAILABILITY_B05}{COMMON}{k} to \gval{TieBlock:Q1_2026:block_end_rank} in Q1. Capture bounds show how much the result can depend on the choice among tied CVEs while keeping list sizes equal.

At the primary budget, selecting earlier publications yields fewer captures than the post hoc uniform expectation in 2025 and more in Q1 (Table~\ref{tab:primary}). This comparison describes the two observed tie blocks; it does not test the tie-breaking rule.

\subsection{Earlier decisions change the evaluation target}
Moving from D+\param{sensitivity_horizon} to D+\param{primary_horizon} changes the available signals and the evaluation target: early KEV entries become ineligible, while the later window admits entries 188--210 days after NVD publication. Each horizon therefore needs its own eligibility assessment and labels.

EPSS captures more early entries in Q1, whereas the strategies capture the same number in 2025. Neither subgroup represents the full comparison. KEV records listing rather than exploitation onset, so these counts do not measure how quickly either score responds to exploitation.

Among CVEs eligible at both horizons, CVSS lists have higher turnover but EPSS has larger 90th-percentile rank movements. Lower list turnover at one budget thus does not imply less movement throughout the ranking.

\subsection{Score provenance and NVD availability}
Most chosen CVSS scores come from non-NVD providers. Disabling those scores changes capture despite using the same nominal score family.

This distinction is relevant to NVD's changing operations. NIST stated that it would no longer routinely provide a separate CVSS score when a CNA score was available, while clarifying that KEV entries had already received priority~\cite{nist2026}. Our data do not establish a policy effect or trend; the cohorts also differ in period, composition, and reconstruction.

\section{Limitations and reproducibility}
\subsection{Outcome, population, and statistical scope}
Absence from KEV does not establish absence of exploitation. The \texttt{dateAdded} field records the listing date, not exploitation onset~\cite{cisakev}. Our \param{outcome_days}-day listing outcome differs from EPSS's 30-day exploitation target; recall and yield measure neither probability calibration nor direct exploit detection.

The cohorts differ in period, composition, and positive counts. Earlier exploration prevents treating all of Q1 as an untouched holdout, although the extension plan preceded its acquisition. We report paired descriptive comparisons without inferential tests, as potential dependence among CVEs has not been modeled. Both cohorts use the same recorded EPSS model version; their populations and score availability differ.

\subsection{Historical reconstruction and corrections}\label{sec:corrections}
Later acquisition cannot guarantee a complete history. We corrected score-provider mapping and the handling of NVD rejection and reinstatement events. Correcting the 2025 initial state added two eligible CVEs at D+\param{primary_horizon} and changed some list memberships, but left primary captured-positive counts unchanged.

For \source{c4_discrepancies} CVEs, current NVD records and the historical reconstruction contain different CVSS vectors. No event dates the change; neither general updates nor \texttt{lastModified} establish when it occurred. Alternative-value and unusable-score scenarios preserve selected lists and captures, but change some complete-case populations and rank statistics. The NVD-availability indicator was held fixed. Rankings use historical values supported by recorded events, not later snapshots.

\subsection{Materials and verification}\label{sec:verification}
Retained inputs, configurations, decision rows, and rankings are identified by hashes. The publication generator exports saved rankings without reranking. Independent recomputation checks processing using the same historical sources. A retained sample of \source{epss_api_pairs} CVE/date pairs matched FIRST API scores exactly.

The first author reviewed all \val{Q1_2026}{D30_AVAILABILITY_B05}{COMMON}{P} eligible Q1 D+\param{primary_horizon} positives, including those selected by neither strategy. This review does not cover all negatives or all 2025 positives. Processing errors are corrected for every affected record, with both strategies and affected analyses recomputed without changing selection rules.

Codex (OpenAI) and Claude (Anthropic) supported code development, testing, and manuscript preparation under the first author's direction and critical review. Methodological decisions remained with the authors, who take responsibility for the methods, results, and manuscript.

The ancillary files provide full metrics for the planned comparisons and post hoc diagnostics, together with records for every primary positive: CVE identifier, reconstructed scores, CVSS source class, ranks, and selection status. Their README maps files to results and documents fields and review scope. Full decision rows, saved rankings, raw inputs, and code are retained but not included in this release.

Q1 data were acquired through an organizational TLS proxy with certificate and hostname verification enabled. Upstream Git blob hashes corroborated EPSS file identity. We compared NVD and KEV record counts with source totals (\texttt{totalResults} for NVD); these checks do not establish content identity. Private environment records are not distributed.
\section{Conclusions}
At a fixed budget, comparing CVSS and EPSS requires an explicit decision cutoff, a common eligible population, documented score availability and provenance, and a stated tie-breaking rule. At D+\param{primary_horizon} and a \param{primary_budget_pct}\% budget, EPSS captured \val{2025}{D30_AVAILABILITY_B05}{EPSS_OPERATIONAL}{captured} of \val{2025}{D30_AVAILABILITY_B05}{COMMON}{P} subsequent KEV positives in 2025, compared with \val{2025}{D30_AVAILABILITY_B05}{CVSS_OPERATIONAL}{captured} for CVSS. Even the most favorable selection within the CVSS cutoff tie block could capture only \val{2025}{D30_AVAILABILITY_B05}{CVSS_OPERATIONAL}{maximum}. In Q1 2026, CVSS captured \val{Q1_2026}{D30_AVAILABILITY_B05}{CVSS_OPERATIONAL}{captured} of \val{Q1_2026}{D30_AVAILABILITY_B05}{COMMON}{P} positives and EPSS captured \val{Q1_2026}{D30_AVAILABILITY_B05}{EPSS_OPERATIONAL}{captured}, but alternative selections within the CVSS block could reverse that ordering. Neither cohort provides a general verdict on the strategies.

Both primary CVSS rankings cut through large blocks at score \tie{2025}{D30_AVAILABILITY_B05}{CVSS_OPERATIONAL}{cutoff_score}. The score identifies the block; the tie-breaking rule determines which CVEs fill the remaining places. Those places make up most of each primary CVSS list. Non-NVD providers supplied most chosen CVSS scores, and disabling those scores changed capture in both cohorts. More than half of subsequent KEV positives remained outside both selected lists in each cohort.

For teams with fixed capacity, three practices follow. State the tie-breaking rule alongside the score used for ranking. Record who assigned each CVSS score. Interpret any single-period comparison, including this one, with its decision cutoff, population, budget, outcome window, and capture bounds (Table~\ref{tab:method}). Asset context, existing catalog evidence, and action costs remain additional inputs to operational decisions.

Automated remediation using score-based rankings likewise needs explicit tie-breaking and budget rules. A further cohort, with inference specified before its results are examined, can test whether the observed patterns carry forward.

\FloatBarrier
\setlength{\bibsep}{0pt}
\renewcommand{\bibfont}{\bibliofont\fontsize{9}{11}\selectfont}
\bibliographystyle{ACM-Reference-Format}
\bibliography{references}
\end{document}